\documentclass[prb%
 reprint,
superscriptaddress,
 amsmath,amssymb,
 aps,floatfix,twocolumn,final,prl,showpacs
]{revtex4-2}

\usepackage{graphicx}
\usepackage{dcolumn}
\usepackage{bm}
\usepackage{textcomp,mathcomp}
\usepackage{verbatim}
\usepackage{physics}
\usepackage{url}

\usepackage{hhline}
\usepackage{tabularx}
\usepackage{tablefootnote}
\usepackage{gensymb}

\usepackage[english]{babel}
\usepackage[table]{xcolor}
\definecolor{lightgray}{gray}{0.9}
\usepackage{epsfig}
\usepackage{graphicx}
\usepackage[utf8]{inputenc}
\usepackage[T1]{fontenc}
\usepackage{scalefnt}
\usepackage{multirow}
\usepackage{float}
\usepackage{amsmath}
\usepackage{flafter}

\begin{document}

\preprint{APS/123-QED}

\title{Magnetic polaron formation in EuZn\textsubscript{2}P\textsubscript{2}}

\author{Matthew S. Cook}
 \affiliation{Los Alamos National Laboratory, Los Alamos, New Mexico 87545, USA}%
\author{Elizabeth A. Peterson}
\affiliation{Los Alamos National Laboratory, Los Alamos, New Mexico 87545, USA}%
\author{Caitlin S. Kengle}
\affiliation{Los Alamos National Laboratory, Los Alamos, New Mexico 87545, USA}%
\author{E. R. Kennedy}
\affiliation{Los Alamos National Laboratory, Los Alamos, New Mexico 87545, USA}%
\author{J. Sheeran}
\affiliation{Los Alamos National Laboratory, Los Alamos, New Mexico 87545, USA}%
\author{Cl\'ement Girod}
\affiliation{Los Alamos National Laboratory, Los Alamos, New Mexico 87545, USA} 
\author{G.S. Freitas}
\affiliation{Instituto de F\'{i}sica ``Gleb Wataghin'', Universidade Estadual de Campinas, Campinas SP 13083-859, Brazil}%
\author{Samuel M. Greer}
\affiliation{Los Alamos National Laboratory, Los Alamos, New Mexico 87545, USA}%
\author{Peter Abbamonte}
\affiliation{Department of Physics, University of Illinois, Urbana, Illinois 61801, USA}
\affiliation{Materials Research Laboratory, University of Illinois, Urbana, Illinois 61801, USA}
\author{P.G. Pagliuso}
\affiliation{Instituto de F\'{i}sica ``Gleb Wataghin'', Universidade Estadual de Campinas, Campinas SP 13083-859, Brazil}%
\author{J. D. Thompson}
\affiliation{Los Alamos National Laboratory, Los Alamos, New Mexico 87545, USA}%
\author{Sean M. Thomas}
\affiliation{Los Alamos National Laboratory, Los Alamos, New Mexico 87545, USA}%
\author{P. F. S. Rosa}
\affiliation{Los Alamos National Laboratory, Los Alamos, New Mexico 87545, USA}%

\date{\today}

\begin{abstract}
Colossal magnetoresistance (CMR) has been observed across many Eu\textsuperscript{2+}-based materials; however, its origin is not completely understood. Here we investigate the antiferromagnetic insulator EuZn\textsubscript{2}P\textsubscript{2} through single crystal x-ray diffraction, transmission electron microscopy, electrical transport, magnetization, dilatometry, and electron spin resonance measurements complemented by density functional theory calculations. Our electrical resistivity data reveal a large negative magnetoresistance, $MR = [R(H)-R(0)]/R(0)$, that reaches $MR = -99.7$\% at 9~T near the antiferromagnetic ordering temperature $T_N=23$ K. Dilatometry measurements show an accompanying field-induced lattice strain. Additionally, Eu\textsuperscript{2+} electron spin resonance reveals a strong ferromagnetic exchange interaction between Eu\textsuperscript{2+} and conduction electrons. Our experimental results in EuZn\textsubscript{2}P\textsubscript{2} are consistent with a magnetic polaron scenario and suggest magnetic polaron formation as a prevailing explanation of CMR in Eu\textsuperscript{2+}-based compounds.

\end{abstract}

\maketitle


\section{\label{sec:level1}I. Introduction\protect\\}

Materials that exhibit colossal magnetoresistance (CMR) have been intensely studied over the past several decades while motivated by fundamental questions on the interplay between charge, spin, orbital, and lattice degrees of freedom and by potential applications in spintronics and data storage \cite{qiu_spin_2018,prinz_magnetoelectronics_1999,kobayashi_room-temperature_1998}. CMR was first observed in mixed valence perovskite manganites and was initially attributed to a double exchange interaction between Mn\textsuperscript{3+} and Mn\textsuperscript{4+} ions \cite{jonker_ferromagnetic_1950,salamon_physics_2001}. CMR has since been observed in many other families of materials that do not display mixed valence, namely the pyrochlores (e.g. Tl\textsubscript{2}Mn\textsubscript{2}O\textsubscript{7}), spinels ($A$Cr\textsubscript{2}Ch\textsubscript{4}, where $A =$ Fe, Cd, Cu, and Ch is a chalcogen), and several distinct Eu-based materials \cite{ramirez_colossal_1997,sinjukow_metal-insulator_2003,wang_colossal_2021,rosa_colossal_2020}. The observed CMR response in several low-carrier density Eu-based compounds has been recently argued to stem from electronic phase separation due to the formation of bound magnetic polarons (MP) -- polarized clusters formed by the the localization of charge carriers due to a strong exchange interaction between the spin of conduction electrons and localized spins of magnetic ions in the lattice \cite{kagan_electronic_2021,calderon_evaluation_2004,storchak_magnetic_2010,ale_crivillero_magnetic_2023,batista_ferromagnetic_2000,dagotto_colossal_2001,meskine_does_2004,arbuzova_nonuniform_2016,solin_conversion_2016}. The coupling between carrier and local moment tends to align the moments in the vicinity of the carrier, which induces an exchange trapping potential \cite{fisk_magnetic_1979,torrance_bound_1972,kuivalainen_magnetic_2009}.

 Typically, MP formation relies on two requirements: strong exchange coupling and a low carrier density. As a result, many Eu$^{2+}$-based magnetic semiconductors have been argued to promote polaron formation, including EuO, EuSe, EuB\textsubscript{6}, EuBiTe\textsubscript{3}, and more recently Eu\textsubscript{5}In\textsubscript{2}Sb\textsubscript{6} \cite{torrance_bound_1972, kuivalainen_bound_1981,umehara_possible_1996,calderon_evaluation_2004,shon_magnetic_2019,ale_crivillero_magnetic_2023}. The existence of magnetic polarons in EuB\textsubscript{6} is also supported by scanning tunneling spectroscopy measurements, wherein clear magnetic inhomogeneities are observed in the paramagnetic state in the electronic density of states \cite{pohlit_evidence_2018}.  

Recently, renewed interest on the low-carrier-density family of trigonal materials EuT\textsubscript{2}X\textsubscript{2} ($T =$ Zn, Cd, Mn and $X =$ P, As, Sb, In) has emerged due to the predicted non-trivial topology both in the paramagnetic and antiferromagnetic states \cite{ma_spin_2019,riberolles_magnetic_2021}. Notably, many of the members of this family also display large negative magnetoresistance, but whether the observed CMR behavior is related to MP formation remains elusive \cite{goforth_magnetic_2008, chen_negative_2020, blawat_unusual_2022,usachov_magnetism_2024, singh_large_2024}. Whether CMR behavior arises out of a semimetallic or insulating band structure has also been debated. 
EuCd\textsubscript{2}As\textsubscript{2}, grown by the tin-flux technique, was initially widely accepted as a topological semimetal, but recent investigations highlight the subtle role of disorder and reveal a clear semiconducting gap of 0.7~eV\cite{santos_magnetic, nelson_revealing_2024}. EuZn\textsubscript{2}P\textsubscript{2}, also grown through the tin-flux technique, shows narrow band-gap behavior in electrical transport measurements, which suggests a more robust semiconducting ground state. Experimentally determining the precise magnitude of the semiconducting gap, however, is challenging \cite{berry_-type_2022, krebber_colossal_2023}. Infrared spectroscopy measurements at room temperature, in conjunction with density functional theory (DFT) calculations, suggest a direct bandgap of 340~meV and an indirect bandgap of 200~meV. EuZn\textsubscript{2}P\textsubscript{2} orders antiferromagnetically at $T_{N}=23$~K, and electrical transport measurements reveal CMR behavior wherein the magnetoresistance, $MR = R(H)-R(0)/R(0)$, reaches -99.9\% at 11~T at the N\'eel temperature. Though magnetic polaron formation was mentioned as a possible explanation of the CMR response \cite{krebber_colossal_2023}, a comprehensive investigation of this scenario is missing.

{\renewcommand{\arraystretch}{1.1}
\begin{table}[]
\caption{Details of crystal data and refinement for EuZn$_2$P$_2$.}
\label{tab:refinement_parameters}

\begin{tabular}{ll}
\hhline{==}
Formula & EuZn$_2$P$_2$
\\ \hline
Space Group  & P$\bar{3}$m1
 \\
 Crystal Size (mm)  & $0.063 \times 0.053 \times 0.008$ \\
a (\r{A}$$)                                                                           & 4.0858(3)  \\
c (\r{A}$$)                                                                           & 7.0056(9)  \\
V (\r{A}$^{3}$)                        & 101.28(2)  \\
Z                                                                                            & 1          \\
T (K)                                                                                        & 293        \\
$\theta$ ($ \degree $)                                                                                     & 2.9 - 30.5 \\
$\mu$ (mm$^{-1}$)                                                                            & 27.59      \\
Measured/Independent Reflections                                                            & 3781/151       \\
R$_{int}$                                                                                    &  0.0563          \\
Extinction Coefficient                                                                       & 0.357(16)  \\
\hhline{==}
\end{tabular}
\end{table}
}

{\renewcommand{\arraystretch}{1.1}
\begin{table}[]
\caption{Refinement results with and without letting the zinc site occupation vary}
\label{tab:refinements}
\begin{tabular}{lll}
\hhline{===}
 & Fixed & Varied \\ \hline
Refinement Parameters    & 10  & 11      \\
$R (F > 2\sigma(F)$)\footnotemark    & 0.0149    & 0.0148\\
$wR( F^2)$\footnotemark & 0.0377   &   0.0370 \\ 
GooF\footnotemark & 1.267 & 1.246 \\ 
Zn occupancy & 1 & 0.992(3) \\
\hhline{===}
\end{tabular}
\footnotetext[1]{$R(F) = \Sigma ||F_0|-|F_c||/\Sigma |F_0|$}
\footnotetext[2]{$ wR(F^2) = \bigl[\Sigma w(F_0^2 - F_c^2)^2/ \Sigma(F_0^2) \bigr]^{1/2} $}
\footnotetext[3]{ $GooF = \bigl( \bigl[ \Sigma w (F_0^2F_c^2)^2 \bigr]/(N_{\mathrm{ref}}- N_{\mathrm{param}}\bigr)^{1/2}$ for all reflections.}
\end{table}
}

Here, we investigate EuZn\textsubscript{2}P\textsubscript{2} grown using indium flux and provide evidence for the formation of magnetic polarons. We perform a comprehensive set of electrical transport measurements that include magnetoresistance and resistivity as a function of applied mechanical stress. The observed CMR is consistent with other Eu-based systems that form magnetic polarons, where the magnetoresistance peaks at or in proximity of the magnetic ordering temperature.  Our magnetization measurements confirm the magnetic anisotropy in Ref.~\cite{krebber_colossal_2023} and show a deviation from Curie-Weiss behavior below $T^{*} = 150$~K, which is consistent with MP formation in the paramagnetic state below $T^{*}$. Using dilatometry, we track the magnetic transition in field and observe field-dependent magnetoelastic coupling. Magnetostriction measurements show an accompanying colossal length change induced by field just above the magnetic transition. Finally, electron spin resonance (ESR) measurements reveal a large ferromagnetic exchange coupling between Eu $4f$ electrons and conduction electrons and suggest that coherent spin-flip scattering develops for $T<200\ K$. Our experimental results support magnetic polaron formation within EuZn\textsubscript{2}P\textsubscript{2} in analogy to other Eu-based magnetic materials, which indicate that magnetic polarons may be ubiquitous in these materials. 

\begin{figure}[!ht]
    \begin{center}
    \includegraphics[width=\columnwidth]{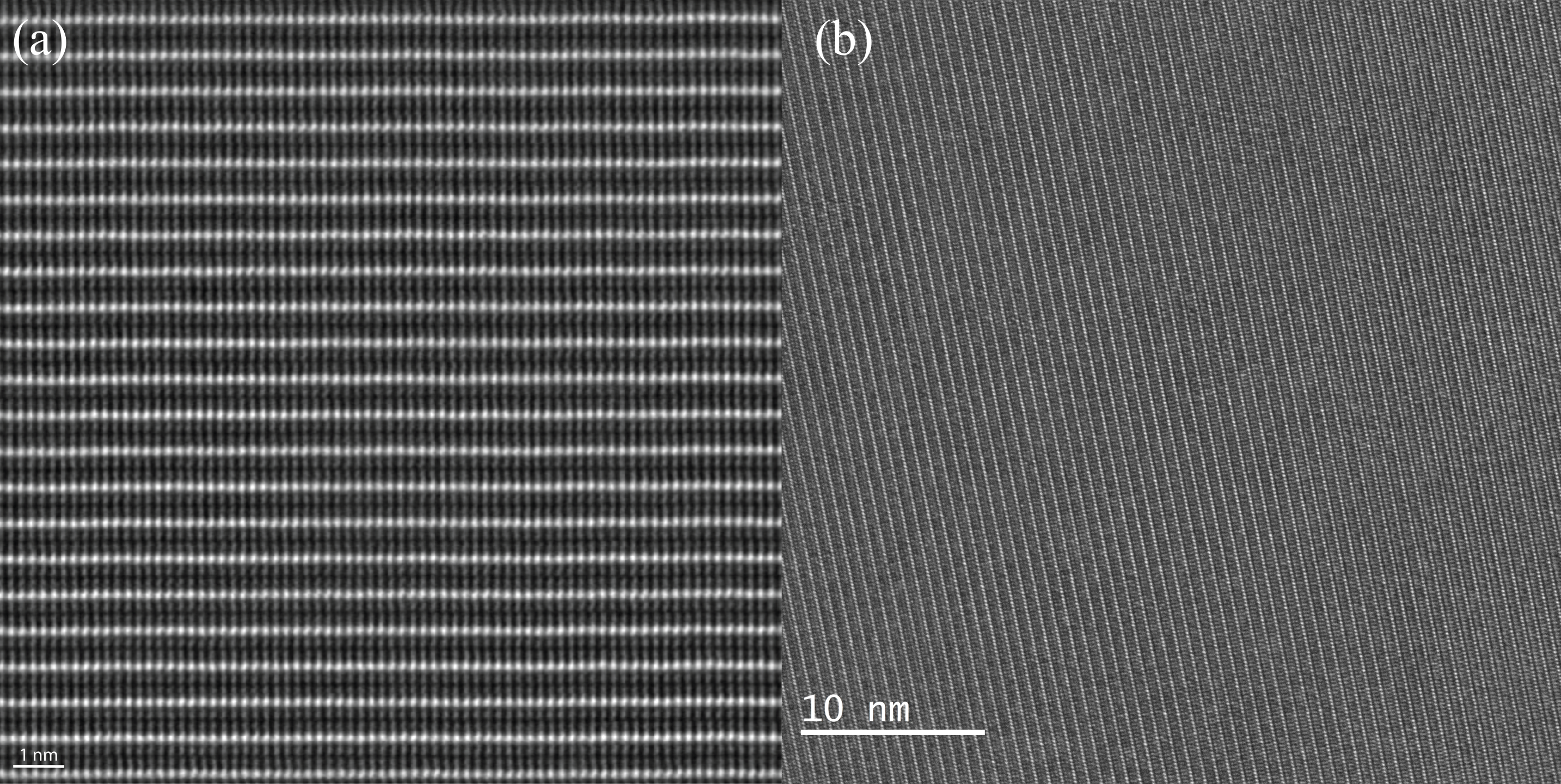}
    \end{center}
    \caption{High-resolution (a) STEM and (b) bright-field TEM images taken along the <10$\bar{1}$0> direction, showing the layered structure of EuZn2P2. In (a), the STEM image reveals bright Eu atoms, while Zn and P atoms are positioned between the Eu planes, but offset from each other. Similarly, in (b), the TEM image highlights the layered structure over a larger region. No defects or vacancies are visible within the structure.}\label{fig:TEM}
\end{figure}


\section{II. Results}

\textbf{Crystal Structure} The results of single crystal x-ray refinements are shown in Tables \ref{tab:refinement_parameters} and \ref{tab:refinements} in the main text, with additional details provided in Tables~\ref{tab:atomic_sites_Zn_fixed_varied_occ}, \ref{tab:ADPs_fixed_Zn_occ}, and \ref{tab:ADPs_varied_Zn_occ} in Appendix B. The refinement was first performed with fixed site occupancies. The resulting lattice parameters shown in Table~\ref{tab:refinement_parameters} are consistent with those reported in the literature for Sn-grown crystals \cite{klufers_ab_1980, berry_-type_2022,krebber_colossal_2023}. The fractional atomic coordinates and isotropic displacement parameters are listed in Table \ref{tab:atomic_sites_Zn_fixed_varied_occ} in Appendix B. The refinement was then repeated letting the Zn site occupancy vary. The metrics of each refinement and the refined Zn occupancy are shown in Table \ref{tab:refinements}. Details of this procedure are further discussed in Appendix B. Allowing Zn occupancy to vary freely yields a value of 99.2(3)\%. Additionally, there is minimal change in the residuals and Goodness of Fit (GooF) parameters. In summary, our results show that all atomic sites are fully occupied and there are no vacancies or indium substitutions.

Transmission electron microscopy measurements were performed to confirm the absence of interstitial substitutions in our crystals. 
Figure~\ref{fig:TEM}(a) shows a high-resolution scanning transmission electron microscopy (HR-STEM) image of EuZn\textsubscript{2}P\textsubscript{2}, whereas Figure~\ref{fig:TEM}(b) shows a high-resolution TEM image. 
The high-resolution images show the expected layered structure of EuZn\textsubscript{2}P\textsubscript{2}. In the STEM image (a), Eu atoms appear bright, with Zn and P atoms positioned between the layers. The structure appears uniform and ordered across the field of view, with no visible structural defects. Our results do not show evidence for either Zn vacancies or In interstitials in imaging or electron energy loss spectroscopy.

\begin{figure*}[!ht]
    \begin{center}
    \includegraphics[width=\textwidth]{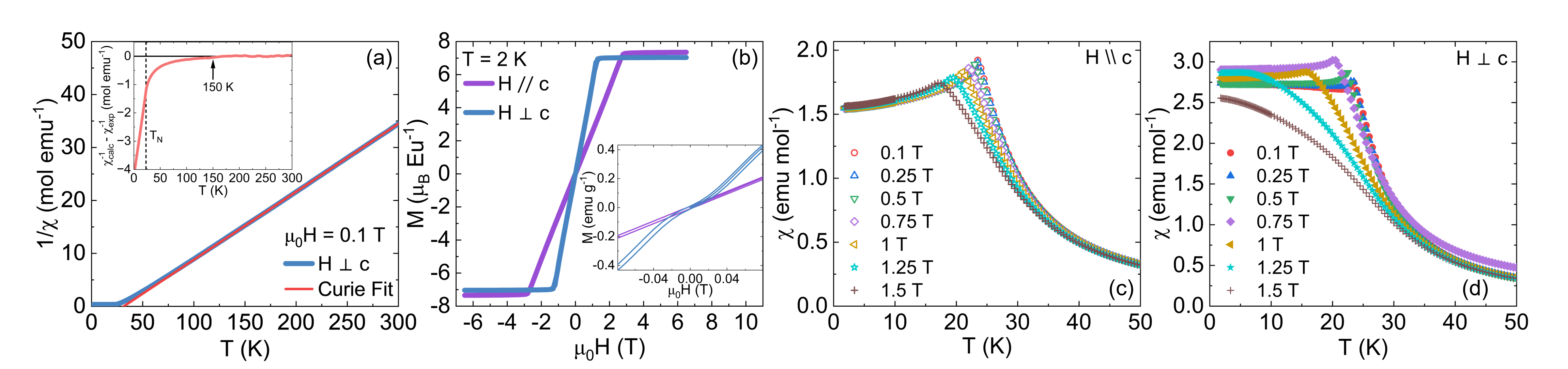}
    \end{center}
    \caption{\label{fig:magnet} (a) Inverse magnetic susceptibility for $\mu_0H=0.1$ T with magnetic field $H\perp\ $c. The Curie-Weiss fit is shown as a solid red line. Inset: Subtraction of experimental data $\chi(T)_{\text{exp}}^{-1}$ from data calculated from Curie-Weiss fit $\chi(T)_{\text{calc}}^{-1}$ that highlights the deviation from Curie-Weiss behavior. (b) Magnetization loops as a function of magnetic field for $H\parallel$~c and $H\perp\ $c. Inset: Weak hysteresis and slope change in the magnetization at low fields for $H\perp\ $c. Magnetic susceptibility as a function of temperature for various magnetic field values with (c) $H\parallel$~c and (d) $H\perp\ $c.}
\end{figure*}


\textbf{Magnetization} We first confirm that the magnetic properties of our In-grown crystals are consistent with those reported for Sn-grown crystals. The temperature dependence of the inverse magnetic susceptibility, $\chi(T)^{-1}$, is presented in Fig.~\ref{fig:magnet}(a) for field applied perpendicular ($H\perp\ $c) to the hexagonal c axis. The inset of Fig.~\ref{fig:magnet}(a), which plots a subtraction of the experimental data from the Curie-Weiss fit, shows that below $T^{*}=150$~K, $\chi(T)^{-1}$ deviates from the expected linear Curie-Weiss behavior. This is the first indication of the presence of interactions beyond mean field. Such deviation has been previously observed in EuZn\textsubscript{2}P\textsubscript{2} and attributed to the formation of magnetic polarons \cite{krebber_colossal_2023}. A similar deviation was also observed in Eu\textsubscript{5}In\textsubscript{2}Sb\textsubscript{6} wherein the magnetic polaron picture is supported by several experiments \cite{rosa_colossal_2020,ghosh_colossal_2022,ale_crivillero_magnetic_2023,souza_microscopic_2022}. Due to the non-linearity below $T^{*}$, a fit to the Curie-Weiss law was performed from 300~K to 150~K, and the extracted parameters yield an effective moment of $\mu_{\mathrm{eff}}=7.895(2)\ \mu_\mathrm{B}$ and a Weiss temperature $\theta_P=31.40(8)$ K. The effective moment is consistent with that expected for an Eu\textsuperscript{2+} moment (7.94 $\mu_\mathrm{B}$), whereas the ferromagnetic Weiss temperature is similar to that found in previous reports \cite{berry_-type_2022,singh_superexchange_2023}.

Anisotropic field-dependent magnetization data are shown in Fig.~\ref{fig:magnet}(b). At low fields, we note a small slope change at approximately 0.01~T for $H\perp$ c (inset of Fig.~\ref{fig:magnet}(b)). The authors of Ref.~\cite{krebber_colossal_2023} also observed similar behavior within the low-field region for $H\parallel$ [210] and attributed it to AFM domains, which varied between several samples. Supporting evidence for domain reorientation in magnetostriction measurements will be presented in the next section. We also observe a weak magnetization hysteresis for $H\perp$ c, consistent with a small in-plane ferromagnetic component. This component is likely due to the canted structure of the magnetic moments that was previously verified through neutron diffraction measurements \cite{krebber_colossal_2023}. The low-temperature magnetic susceptibility shown in Fig.~\ref{fig:magnet}(c)-(d) highlights the magnetic field dependence of the AFM transitions. The anisotropy and field dependence observed in our measurements are in agreement with Refs. \cite{krebber_colossal_2023,singh_superexchange_2023}.

\begin{figure}[ht]   
\includegraphics[width=0.9\columnwidth]{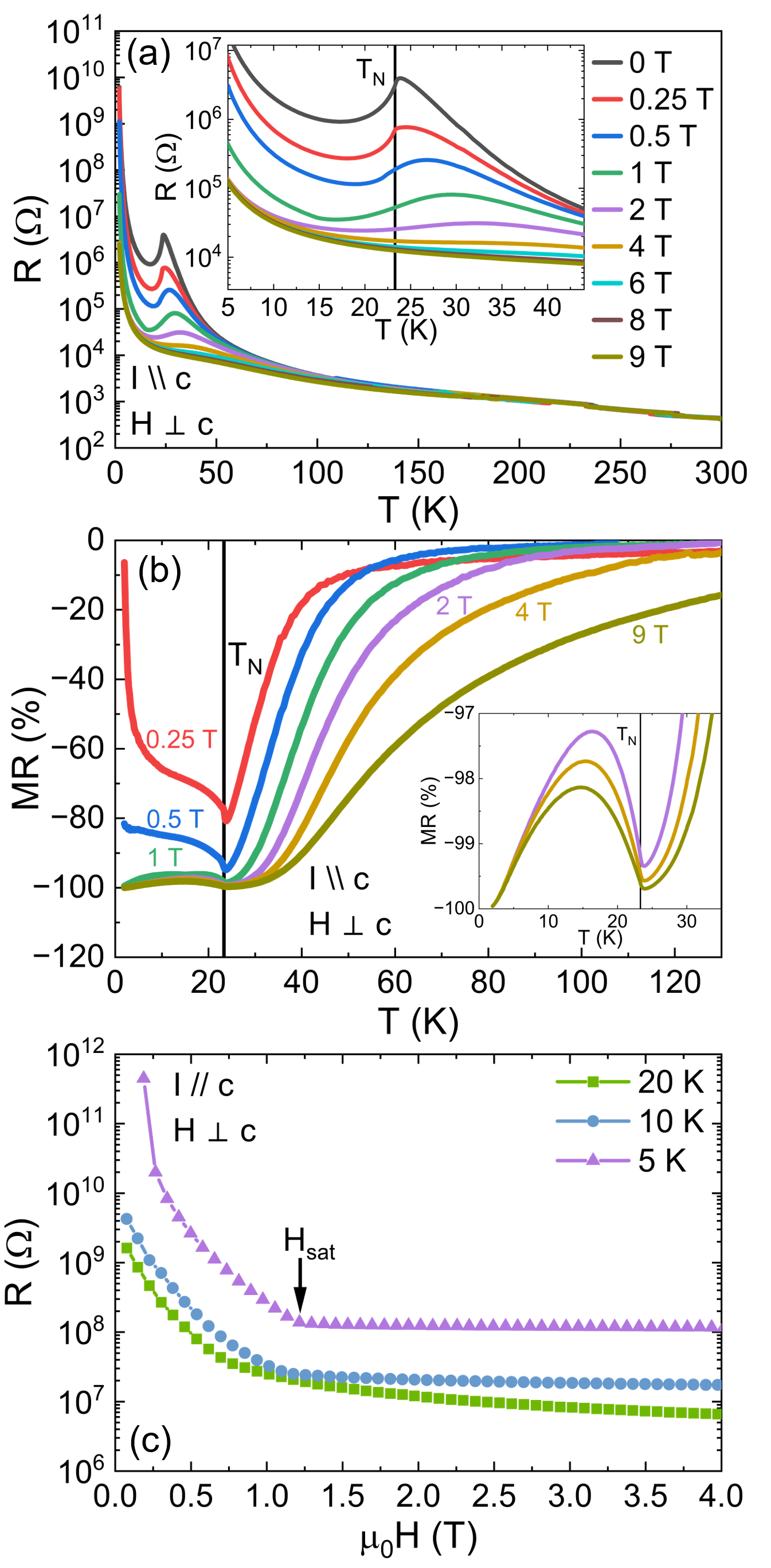}
\vspace{6pt}
\caption{\label{fig:Resist_2panel}(a) Resistance as a function of temperature for various in-plane magnetic fields. The inset shows a close view of the field evolution of the peak in the resistance at $T_N$. (b) Magnetoresistance as a function of temperature at various field values that is calculated from the data in (a). Inset: Close up of low temperature $MR$. (c) Resistance isotherms as a function of field.}
\end{figure}

\textbf{Electrical Transport} We now turn to our attempts to understand magnetic polaron formation in this system through electrical transport measurements, where these quasiparticles are known to dominate or significantly influence transport in other magnetic polaron systems. Figure~\ref{fig:Resist_2panel}(a) shows the temperature dependence of the resistance, $R(T)$, for various in-plane fields to 9~T. The observed temperature dependence differs from that reported by Krebber \textit{et al.} \cite{krebber_colossal_2023} in several aspects. First, we observe no minimum in $R(T)$ at high temperatures. Second, the resistance reported here changes by over 7 orders of magnitude from room temperature to 5~K, whereas the resistivity in Ref.~\cite{krebber_colossal_2023} only changes by 3 orders of magnitude. Other samples from the same growth batch show even larger resistances at low temperatures; however, we were unable to measure these samples to $T_N$ due to current leakage limitations. In indium-grown samples, we also observe a clear peak in the resistance that coincides with $T_N$ in zero field. We attribute varying properties in our samples to a small volume fraction of flux inclusions that create one or multiple parallel conduction paths. Further evidence of this claim will be presented when discussing the experimental transport gap. Further studies are required to establish the origin of the marked differences between In-grown and Sn-grown crystals.

The peak in resistance just above $T_N$ moves to higher temperatures with applied field, whereas $T_N$ itself decreases with field. The application of fields above 2~T results in a rapid suppression of the peak magnitude. Notably, the resistance at $T_N$ decreases by over 2 orders of magnitude in a 9~T field. This behavior is reminiscent of the magnetotransport observed in Eu\textsubscript{5}In\textsubscript{2}Sb\textsubscript{6}, wherein a broad maximum close to the AFM ordering temperature in zero field is pushed to higher temperatures and decreases in magnitude as the field is increased \cite{rosa_colossal_2020}. In a magnetic polaron scenario, the peak in resistivity in zero field is related to the delocalization of charge carriers as the polarons grow in size and begin to overlap. There are, however, two clear differences between the resistive behavior of these two materials. For Eu\textsubscript{5}In\textsubscript{2}Sb\textsubscript{6}, the decrease in resistance below the peak in resistance is much larger than in EuZn\textsubscript{2}P\textsubscript{2}. In addition, the activated behavior in $R(T)$ of Eu\textsubscript{5}In\textsubscript{2}Sb\textsubscript{6} breaks down below $T=40\ K$, the temperature at which polarons start to interact $via$ short-range interactions. We argue that these differences are caused by the differences in the carrier concentration between the two materials. As we will show later, the carrier density in EuZn\textsubscript{2}P\textsubscript{2} is smaller than that in Eu\textsubscript{5}In\textsubscript{2}Sb\textsubscript{6}, which may suppress short-range interactions between smaller polarons. Magnetic polarons are well defined and non-overlapping when $n\xi^3\ll{1}$, where $\xi$ is the magnetic correlation length and $n$ is the carrier density \cite{calderon_evaluation_2004}; thus a smaller carrier density can result in more isolated clusters. 

Figure~\ref{fig:Resist_2panel}(b) plots the magnetoresistance as a function of temperature at various field values. Here, the magnetoresistance is given by $MR=\frac{R(H)-R(0)}{R(0)}\times100\%$. For each field value the $MR$ peaks close to $T_N$. In addition, a large negative magnetoresistance of -99.7\% at 9~T, known as CMR, is observed close to $T_N$ at $T=23.8$~K. This behavior resembles previous observations in Eu-based compounds that display magnetically driven electronic phase separation \cite{das_magnetically_2012}. In a magnetic polaron picture, a temperature-driven percolation effect is expected near $T_N$ in zero field, such that the CMR effect from field-induced delocalization of carriers peaks at this temperature. 

\begin{figure}[ht]   
\includegraphics[width=\columnwidth]{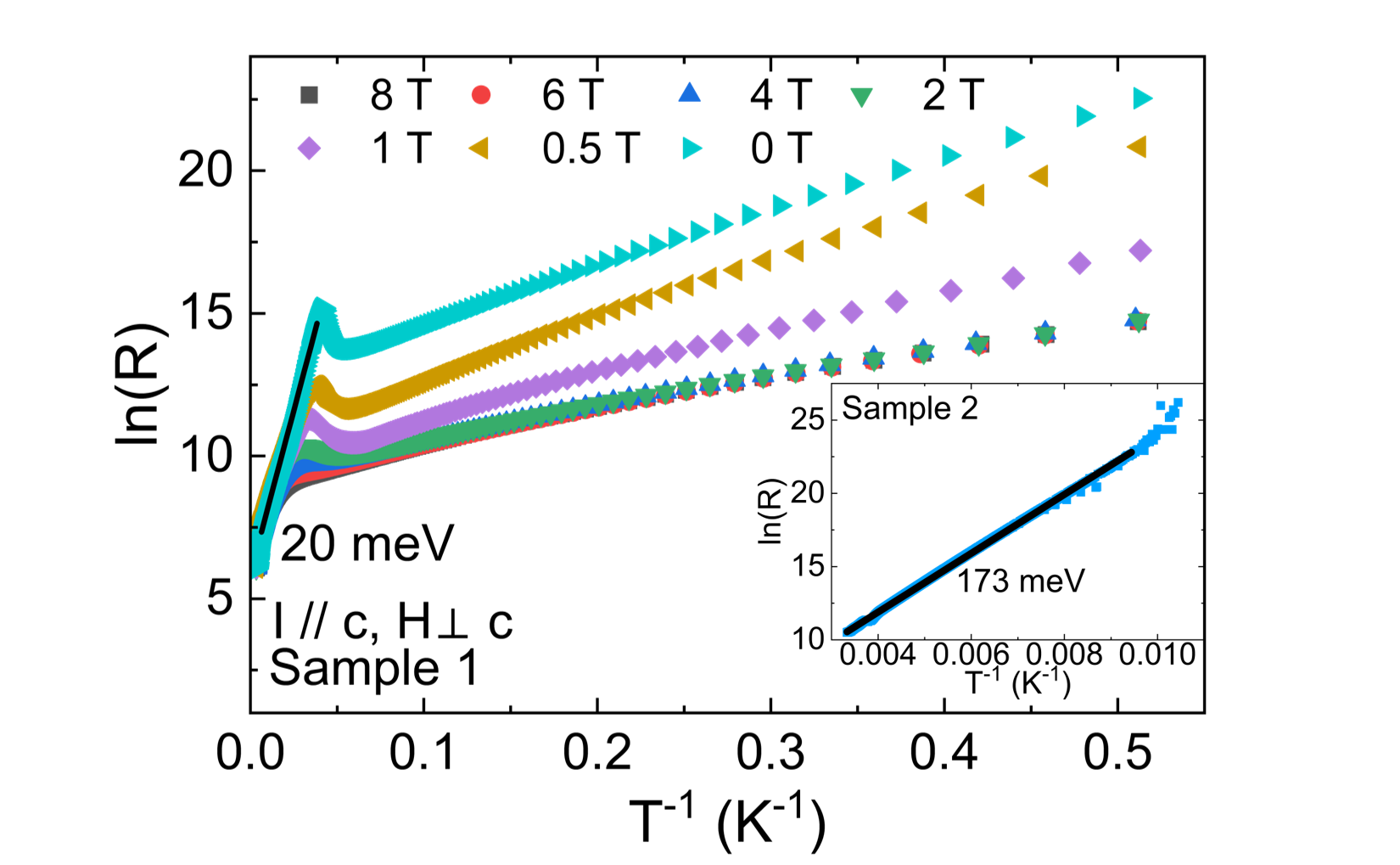}
\vspace{6pt}
\caption{\label{fig:activation} An Arrhenius plot of the resistance data plotted in Fig.~\ref{fig:activation}, noted as Sample 1. The solid black line represents a linear fit to determine the effective energy gap. Inset: An Arrhenius plot of Sample 2 from the same growth batch.}
\end{figure}

The magnetoresistive behavior below the transition is puzzling: the low-temperature MR is also remarkably field dependent, and the MR reaches a maximum value of -99.96\% at $T=2\ $K and $\mu_0H=9\ $T. To explain this behavior, one needs to consider three effects. First, as the field is increased towards the polarized state, there is likely a reduction in magnetic scattering that results in negative magnetoresistance. Another contribution to the negative $MR$ may be the persistence of ferromagnetic clusters within the magnetically ordered state, as was previously proposed for EuCd\textsubscript{2}P\textsubscript{2} \cite{usachov_magnetism_2024}, EuO \cite{snow_magnetic_2001}, the electron doped manganite CaMnO\textsubscript{3-x} \cite{trokiner_magnetic_2009}, as well as Eu\textsubscript{5}In\textsubscript{2}Sb\textsubscript{6} \cite{dawczak-debicki_thermodynamic_2024}. Finally, recent DFT calculations for both EuCd\textsubscript{2}P\textsubscript{2} \cite{usachov_magnetism_2024} and EuCd\textsubscript{2}As\textsubscript{2} under pressure \cite{du_consecutive_2022} in the AFM and FM states show that there is a decrease in the gap size for EuCd\textsubscript{2}P\textsubscript{2} and a full suppression of the insulating state for EuCd\textsubscript{2}As\textsubscript{2} once the moments are fully polarized. Both of these compounds show considerable MR at temperatures well below the magnetic ordering temperature, and field-induced changes in band structure in the ordered state are thought to be a contributing factor. Du et al. point out that the low-temperature $MR$ for EuCd\textsubscript{2}As\textsubscript{2} at 1.5 GPa saturates at the same field values as the magnetization. This also is found in EuZn\textsubscript{2}P\textsubscript{2}, as shown in Fig.~\ref{fig:Resist_2panel}(c), where $H_{\text{sat}}(5\ \text{K})=1.3\ \text{T}$ for field applied perpendicular to c \cite{krebber_colossal_2023}. This suggests a related mechanism for negative $MR$ in the AFM state. As another direct comparison, the temperature dependence of the $MR$ in EuCd\textsubscript{2}P\textsubscript{2} at 5 T shows a very similar trend to EuZn\textsubscript{2}P\textsubscript{2}. As the resistivity was only measured to about 6~K, it is unknown if the $MR$ reaches its maximum value at lower temperatures.

To determine whether a field-induced polarization of Eu$^{2+}$ moments could have an effect on the transport behavior in this system, we perform DFT calculations in both the AFM and FM states. As we will discuss later, there is indeed a reduction of the band-gap by 25\% as one moves from an AFM to a FM state. The associated changes of band structure in a polarizing field may be a contributing factor in the observed MR within the ordered state.


As previous reports on Eu-based semiconductors show varying behavior in electrical transport, it is important to address the effects of sample dependence on transport behavior, and in particular, the activated behavior of electrical conduction. Figure \ref{fig:activation} shows an Arrhenius plot of sample 1 at various fields. The energy gap is determined to be $E_g=20$ meV for 30 K $\leq{T}\leq$ 200 K, which is significantly smaller than the 110 meV gap deduced in Ref. \cite{berry_-type_2022} from electrical resistivity. The inset of Fig. \ref{fig:activation} shows a similar plot of another sample from a separate growth attempt, for which the energy gap is 173 meV in the same temperature range.  

There are two likely scenarios contributing to the variation in the extracted gap. One, there is an impurity band close to the conduction band that is decreasing the effective gap energy. Our single crystal x-ray diffraction and transmission electron microscopy measurements, however, do not support this scenario. It is still possible that very small amounts of defects are present and influence such a narrow-gap semiconductor. However, a more likely scenario is that there are flux inclusions that grow side-by-side with EuZn\textsubscript{2}P\textsubscript{2} and create parallel conduction paths that affect the temperature dependence and the apparent transport gap. In fact, upon polishing the EuZn\textsubscript{2}P\textsubscript{2} single crystal used in the electrical resistivity measurements shown in the main panel of Fig. 4, significant indium flux inclusions were readily apparent. We therefore argue that the semiconducting gap in EuZn\textsubscript{2}P\textsubscript{2} likely lies in the ~200 meV range, as shown in the inset of Fig. 4. Attempts were made to simulate the resistance of a single In conduction path in parallel with EuZn\textsubscript{2}P\textsubscript{2}; however, the correct temperature dependence as well as the magnitude of change in activation energy that we observe between samples could not be reproduced. As there are multiple possible spatial configurations of flux inclusions within these samples, the exact current path is likely complex, which hinders an accurate simulation.

\begin{figure}[ht]
\includegraphics[width=3.375in]{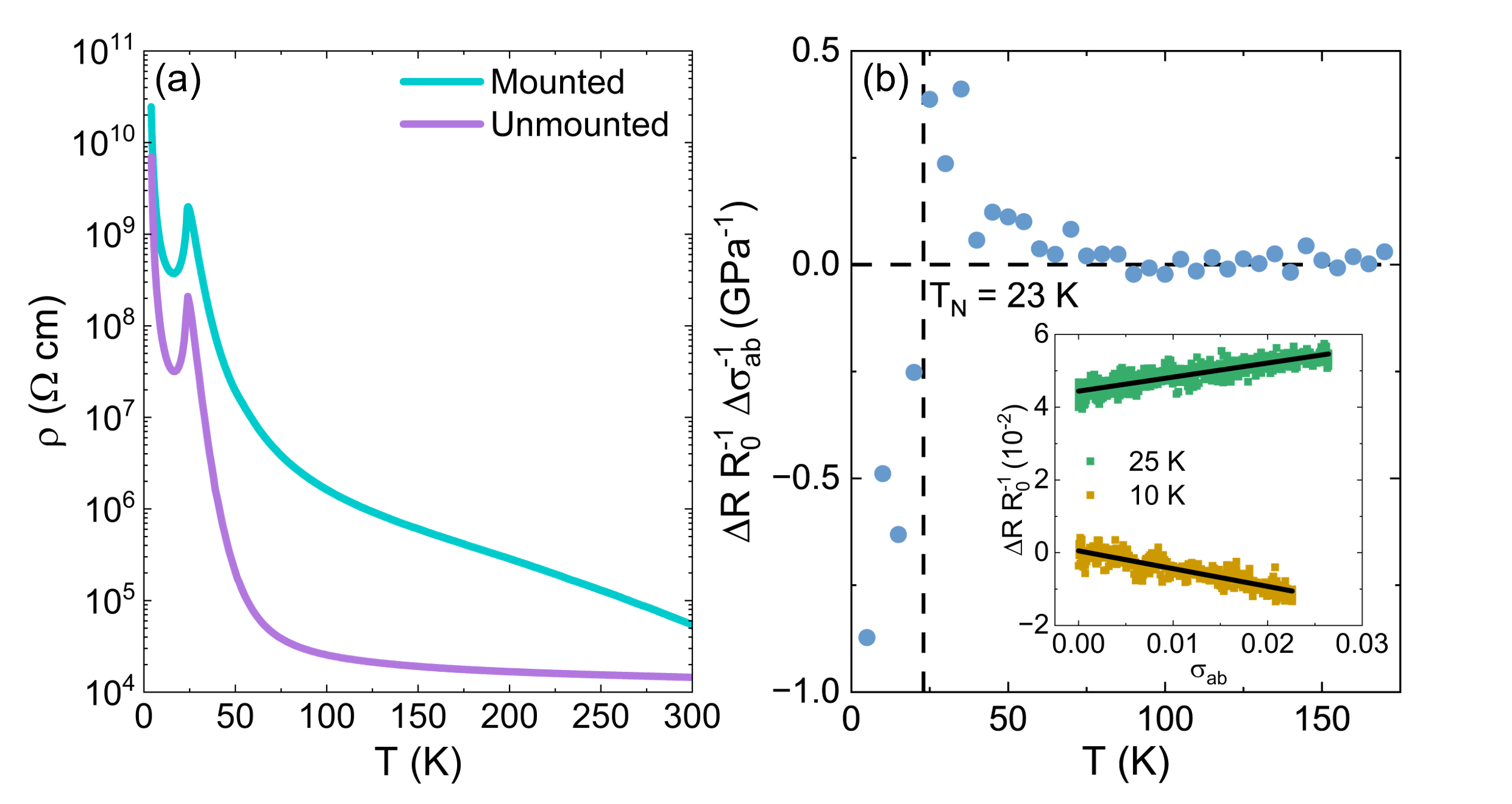}
\caption{\label{fig:stress} (a) Resistivity comparison between a sample mounted within the strain cell (turquoise) and measured on a PPMS puck (violet). (b) The slope of the linear fits to the $\Delta{R}/R_0$ vs $\sigma$ data which represents the temperature dependence of the piezoresistance. Inset: Example linear fits (black lines) of the normalized resistance vs stress applied along the ab-plane. The data for 10 K and 25 K have been offset for visual clarity.}
\end{figure}

As a complimentary consideration, we perform DFT calculations with a random 3\% substitution of In on Zn sites. The results of this analysis are presented in Appendix C. As shown in Fig.~\ref{fig:appendft}, this type of indium defect results in the formation of an impurity band, which at 3\% concentration would result in metallic conduction. This is presented as a proof of principle as an alternative explanation for the observed sample dependence, but we highlight that we did not find any experimental evidence of excess In with single crystal x-ray diffraction or energy dispersive x-ray spectroscopy measurements.


\begin{figure}[!hb]
\includegraphics[width=3.375in]{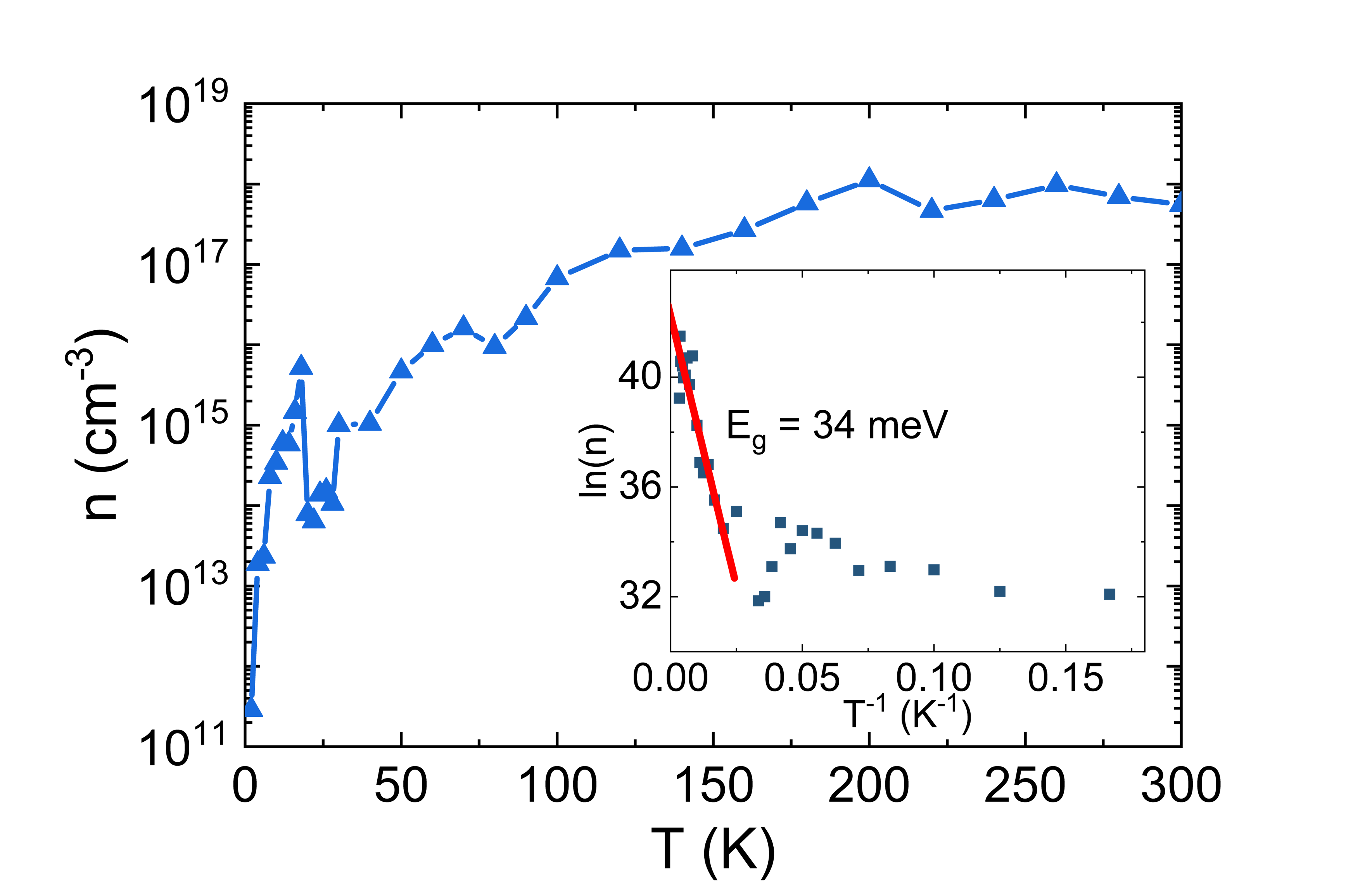}
\caption{\label{fig:carrier_concentration} The carrier concentration of EuZn\textsubscript{2}P\textsubscript{2}, deduced from AC Hall effect measurements.}
\end{figure}

\begin{figure}[ht]
    \centering
    \includegraphics[width=1.0\linewidth]{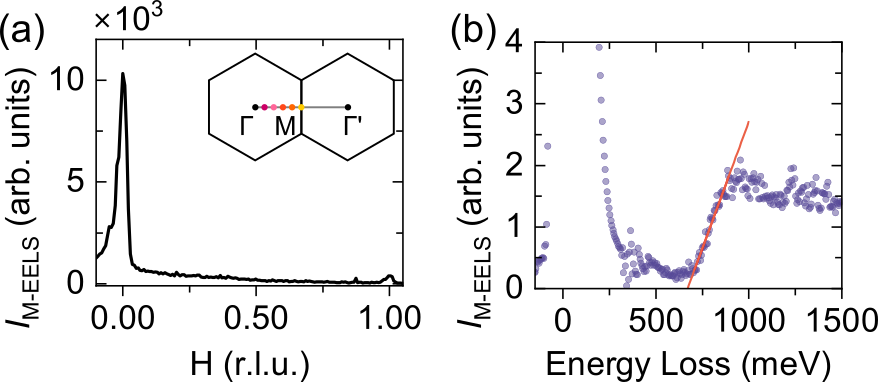}
    \caption{(a) M-EELS momentum scan along ($H,0$) showing Bragg peaks with FWHM of 0.035 r.l.u. (0.0216 {\r{A}}$^{-1}$. (b) $q=0$ M-EELS spectrum with evidence of interband transitions at $\sim600$ meV. The line represents a linear fit used to determine the value of the bandgap.}
    \label{fig:gap_resolution}
\end{figure}

\begin{figure*}
    \centering
    \includegraphics[width=1\linewidth]{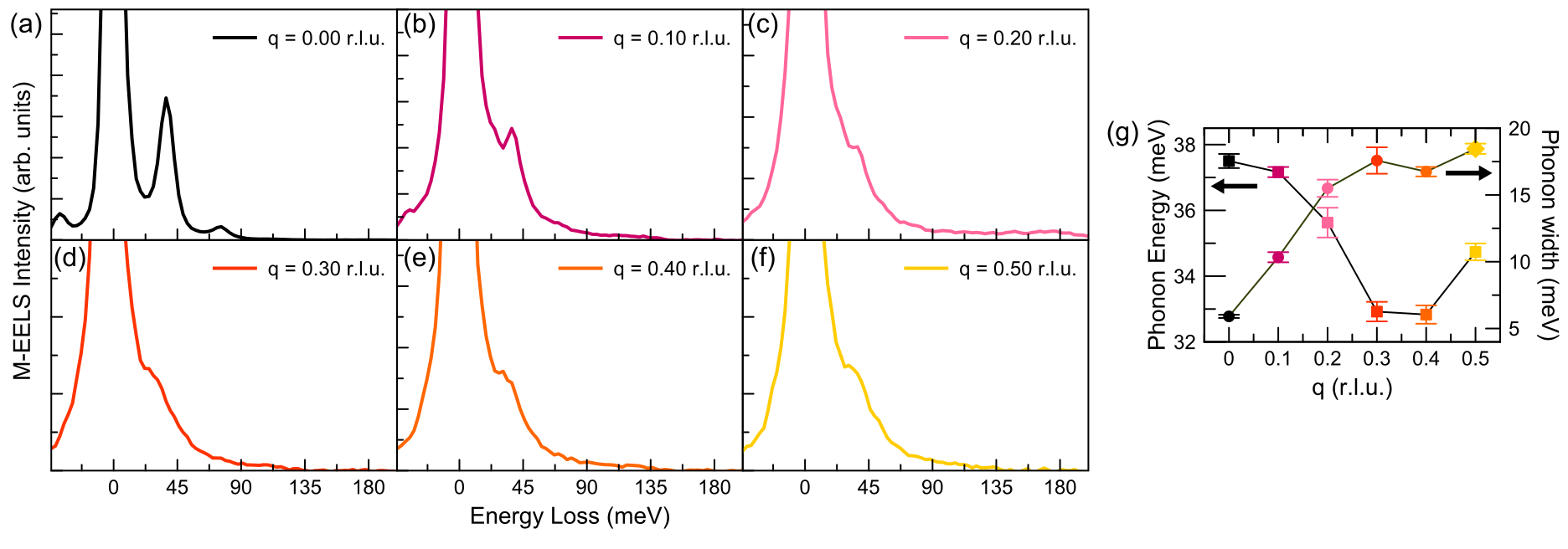}
    \caption{Low-energy M-EELS spectra at (a) $q= 0$, (b) $q=0.1$ r.l.u., (c) $0.2$ r.l.u., (d) $0.3$ r.l.u., (e) $0.4$ r.l.u., and (f) $0.5$ r.l.u. (g) Fit results for the energy and width of the low energy optical phonon.}
    \label{fig:phononfitting}
\end{figure*}

We now turn to the investigation of electrical resistance under uniaxial pressure. Recent studies on EuZn$_2$P$_2$ under large hydrostatic pressures report metallic resistivity at 18.7 GPa in Sn grown samples, with a significant enhancement of the magnetic ordering temperature as pressure increases \cite{PhysRevB.110.014421}. In Eu$_5$In$_2$Sb$_6$, a strong connection between the colossal magnetoresistance and piezoresistance was observed and attributed to the presence of anisotropic MP \cite{ghosh_colossal_2022}. To test for a similar relationship here, we measure the response of resistivity to in-plane uniaxial pressure. In this measurement scheme, we define positive stress as compressive. The resistivity of the sample was measured before and after mounting to the pressure cell, as shown in Fig.~\ref{fig:stress}(a). There is a significant difference in the resistivity of the mounted sample that may be a result of the strain induced by gluing the sample with stycast to the stress cell. This result indicates that EuZn\textsubscript{2}P\textsubscript{2} is sensitive to applied stress; however, the direction of stress applied from the mounting procedure is unclear and may include shear forces.

The inset of Fig.~\ref{fig:stress}(b) shows the dependence of the normalized resistivity on uniaxial pressure along the \textit{ab} plane for $T=10$ K and 25 K. We fit a line to the raw data for each temperature and plot the slope in the main panel of Fig.~\ref{fig:stress}(b), which reveals the temperature dependence of the piezoresistance coefficient. Unexpectedly, there is a small piezoresistance effect near $T_N$ (dashed vertical line). The resistance change close to the magnetic transition appears due to a slight shift in the ordering temperature with applied uniaxial pressure. As the peak shifts slightly higher in temperature, there is a sign change in the piezoresistance from positive to negative at $T_N$. This lack of response is in contrast to Eu\textsubscript{5}In\textsubscript{2}Sb\textsubscript{6}, where Ghosh \textit{et al.} observed a colossal (negative) piezoresistance that follows similar trends to the CMR response \cite{ghosh_colossal_2022}. This was attributed to the percolation of magnetic polarons, in analogy with magnetic field. The differences in carrier concentration mentioned above could also explain this discrepancy. In a lower carrier density material, magnetic polarons may be smaller, well separated, and have a shorter localization length. Another possibility is the anisotropy intrinsic to MP formation in materials with anisotropic exchange interactions, as proposed for Eu\textsubscript{5}In\textsubscript{2}Sb\textsubscript{6} \cite{ghosh_colossal_2022, dawczak-debicki_thermodynamic_2024}. Ellipsoidal magnetic polarons may also be present here, such that a larger response may occur along a different direction. It should also be noted that the resistivity of our measured sample in this experiment indicates parallel conduction paths, as the overall change in resistivity and transport gap is quite small. Flux inclusions were later confirmed, as the sample broke along an inclusion boundary when sufficient force was applied. Measuring the piezoresistance in these samples to higher pressures and along multiple directions would be more informative, though finding a suitably sized sample that is free of inclusions has proven challenging.


Finally, to determine the carrier concentration of EuZn\textsubscript{2}P\textsubscript{2}, we employed ac Hall measurements. This type of measurement is useful in estimating the carrier concentration in low mobility or highly resistive materials where the desired Hall voltage is very small compared to the undesired voltage that comes from imperfect contact geometry. A derivation of the Hall coefficient determined using this technique is presented in Appendix D.

Figure~\ref{fig:carrier_concentration} shows a plot of the carrier concentration that is calculated for EuZn\textsubscript{2}P\textsubscript{2} using this method. The temperature dependence of the carrier concentration fits activated behavior at higher temperatures, as shown in the inset of Fig.~\ref{fig:carrier_concentration}. We estimate the energy gap value $E_g=34$~meV by fitting a line to the Arrhenius plot for 50 K $\leq{T}\leq$ 300 K. The carrier concentration at room temperature is given by $5.6\times10^{17}\ \text{cm}^{-3}$, in the typical range for a narrow-gap semiconductor. This is fairly close to an estimate given by Ref. \cite{krebber_colossal_2023} ($1.5\times10^{17}\ \text{cm}^{-3}$), which was found using band-structure calculations in accordance with the thermal excitation of intrinsic carriers. This value is about an order of magnitude smaller than that found for Eu\textsubscript{5}In\textsubscript{2}Sb\textsubscript{6} at 300 K using the same method \cite{unpublished}. This corroborates the scenario described above wherein differences in CMR and piezoresistance between EuZn\textsubscript{2}P\textsubscript{2} and Eu\textsubscript{5}In\textsubscript{2}As\textsubscript{6} stem from differences in carrier density.

\textbf{M-EELS} We used momentum-resolved electron energy loss spectroscopy (M-EELS) described in detail elsewhere \cite{vig2017measurement} to measure samples of EuZn$_2$P$_2$.
M-EELS is sensitive to the imaginary part of the dynamic susceptibility, $\chi^{\prime\prime}$, such that lattice and electronic effects contribute. 
While typically used to probe collective modes, low-$q$ measurements, particularly at $q=0$, are sensitive to single-particle excitations.

M-EELS measurements were performed at $T= 300$ K on samples of EuZn$_2$P$_2$. The large resistivity of the samples at low temperatures disallowed M-EELS measurements below room temperature. High sample surface quality was confirmed by resolution-limited Bragg scattering with FWHM of $0.02$ \r{A}$^{-1}$ shown in Fig.~\ref{fig:gap_resolution}(a). Figure \ref{fig:gap_resolution}(b) shows the M-EELS spectrum at $q=0$ up to 1.5 eV. 
A clear direct band gap onset is visible around 600 meV, consistent with the predictions from DFT shown in Fig.~\ref{fig:DFTgroup}(c). 
The weak and rapidly decreasing contribution of single-particle excitations to $S(q,\omega)$ makes measurement of the smaller indirect band gap at finite $q$ infeasible.

In contrast, high-intensity optical phonons are visible throughout the Brillouin zone, as shown in Fig.~\ref{fig:phononfitting}. 
The $q=0$ spectra shows a phonon at $\omega_{phonon}=37.5$ meV, consistent with observations in previous FTIR measurements \cite{krebber_colossal_2023}. 
There is a multiphonon scattering event at $\omega_{phonon}\times2 = 75$ meV, a phenomenon often observed in M-EELS measurements of strongly ionic semiconductors \cite{PhysRevB.108.205102,PhysRevLett.120.237001}. 
As $q$ is increased, this optical phonon softens, reaching a minimum value of 33 meV. 
Concurrently, the phonon linewidth increases by a factor of $\sim3.5$. 
Due to this softening and broadening at the Brillouin zone edge, the phonon overlaps significantly with the elastic line. 
The elastic line also broadens with $q$, suggestive of an acoustic phonon with slow sound velocity.
Yet, the optical phonon peak remains resolvable from the elastic scattering line, likely due to avoided crossing with an acoustic phonon buried in the tails of the elastic line. 

\begin{figure*}[!ht]
    \begin{center}
    \includegraphics[width=\textwidth]{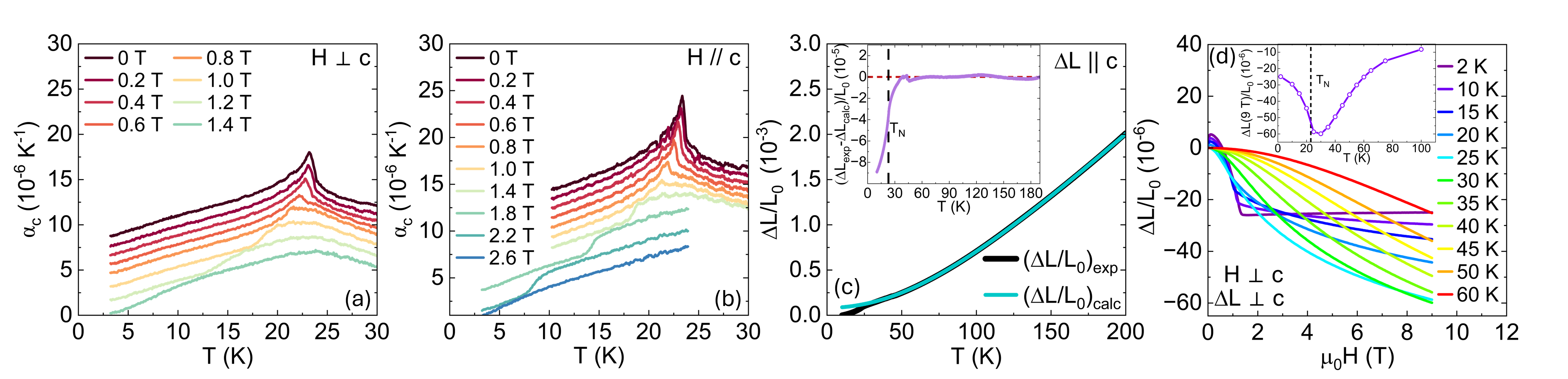}
    \end{center}
    \caption{\label{fig:dil} Thermal expansion coefficient as a function of temperature for various magnetic field values for (a) $H\perp\ $c and (b) $H\parallel$~c which tracks the magnetic phase transition. (c) Relative length change along the c direction as a function of temperature with corresponding fit in light blue. Inset: Difference of calculated relative length change and experimental data. (d) Relative length change along the ab-plane as a function of field for $H\perp$~c at temperatures above and below the magnetic transition. Inset: The magnitude of relative length change at 9 T along the ab-plane.    }
\end{figure*}

\textbf{Dilatometry} The temperature dependence of the thermal expansion coefficient $\alpha=(1/L_0)(d{L}/d{T})$ for several field values is shown in Fig.~\ref{fig:dil}(a) and \ref{fig:dil}(b) with the raw data of each curve shifted upwards for visual clarity. We define $\alpha_c$ as the thermal expansion coefficient along the \textit{c} axis while $\alpha_{ab}$ is measured in the \textit{ab} plane. The magnetic field direction is parallel to that of the measured thermal expansion in both data sets. The anomalies present in the thermal expansion coefficient correspond to the onset of magnetic order in specific heat and magnetic susceptibility measurements. The peaks are very well defined at low fields but broaden significantly as the field is increased. This may be due to the gradual polarization of moments as the field magnitude is increased which reduces the lattice strain associated with the transition.

The relative length change of the sample $\Delta{L}/L_0$ along the c axis shown in Fig.~\ref{fig:dil}(c) can be fit to a semiclassical model that considers a Debye approximation and an electronic contribution to the lattice displacement, as described in Ref.~\cite{MUKHERJEE1994241}. This model is able to describe the data well from approximately $50\ \mathrm{K}\leq{T}\leq{200}\ \mathrm{K}$. The deviation above the magnetic ordering temperature, shown in the inset of Fig.~\ref{fig:dil}(c) as $(\Delta{L}/L_0)_{\mathrm{exp}}~-~(\Delta{L}/L_0)_{\mathrm{calc}}$, is likely from a distortion of the lattice associated with short-range magnetic correlations. Lattice distortions have been long known to accompany magnetic polaron formation in the manganites \cite{ibarra_large_1995,teresa_evidence_1997}. Anomalous lattice strain associated with the formation of magnetic polarons was also reported in Ref. \cite{dawczak-debicki_thermodynamic_2024} for Eu$_5$In$_2$Sb$_6$ within the same energy scale. Similarly, we associate the sizable deviation from the modeled temperature dependence in EuZn$_2$P$_2$ with magnetic polaron effects.

The relative length change of the sample in field $\Delta{L}(H)/L_0$ in the PM and AFM states is plotted in Fig.~\ref{fig:dil}(d). At very low fields in the magnetically ordered state, there is an abrupt feature which we attribute to the reorientation of AFM domains with increasing field. This corresponds well with the slope change observed in the low-field region of $M(H)$ at 2~K for $H\perp{c}$. At 2~K, there is a rapid length contraction associated with the polarization of moments in field, which saturates when the moments are completely polarized. This behavior follows that of the magnetization and MR response. As the temperature increases, the length change is more gradual, but it does not saturate as the moments become polarized. Instead, the magnitude of the $\Delta{L}(H)/L_0$ at high fields smoothly increases through the PM region. To track this behavior, we plot the temperature dependence of $\Delta{L}(H)/L_0$ for $\mu_0H=9$ T, shown in the inset of Fig.~\ref{fig:dil}(d). The largest response is observed slightly above the magnetic ordering temperature, near the same temperature at which the magnitude of CMR also has a local maximum. 

Previous work on ferromagnetic EuB\textsubscript{6} has shown a connection between lattice strain and the field-induced charge carrier delocalization of magnetic polarons \cite{manna_lattice_2014}. The authors observed an anomalous lattice strain in the thermal expansion coefficient that begins at $T\sim$ 35 K, where the formation and stabilization of magnetic polarons is thought to occur. They also compared the observed magnetic contribution of the thermal expansion coefficient in field to the expected theoretical value using a mean-field theory approximation. This comparison also shows that there is a large deviation that is greatest at the temperature where MPs percolate. Finally, similar to our observations, magnetostriction measurements in EuB\textsubscript{6} show the largest relative length change close to the transition temperature, where the CMR effect is the greatest.


\begin{figure*}[ht]
    \begin{center}
    \includegraphics[width=\textwidth]{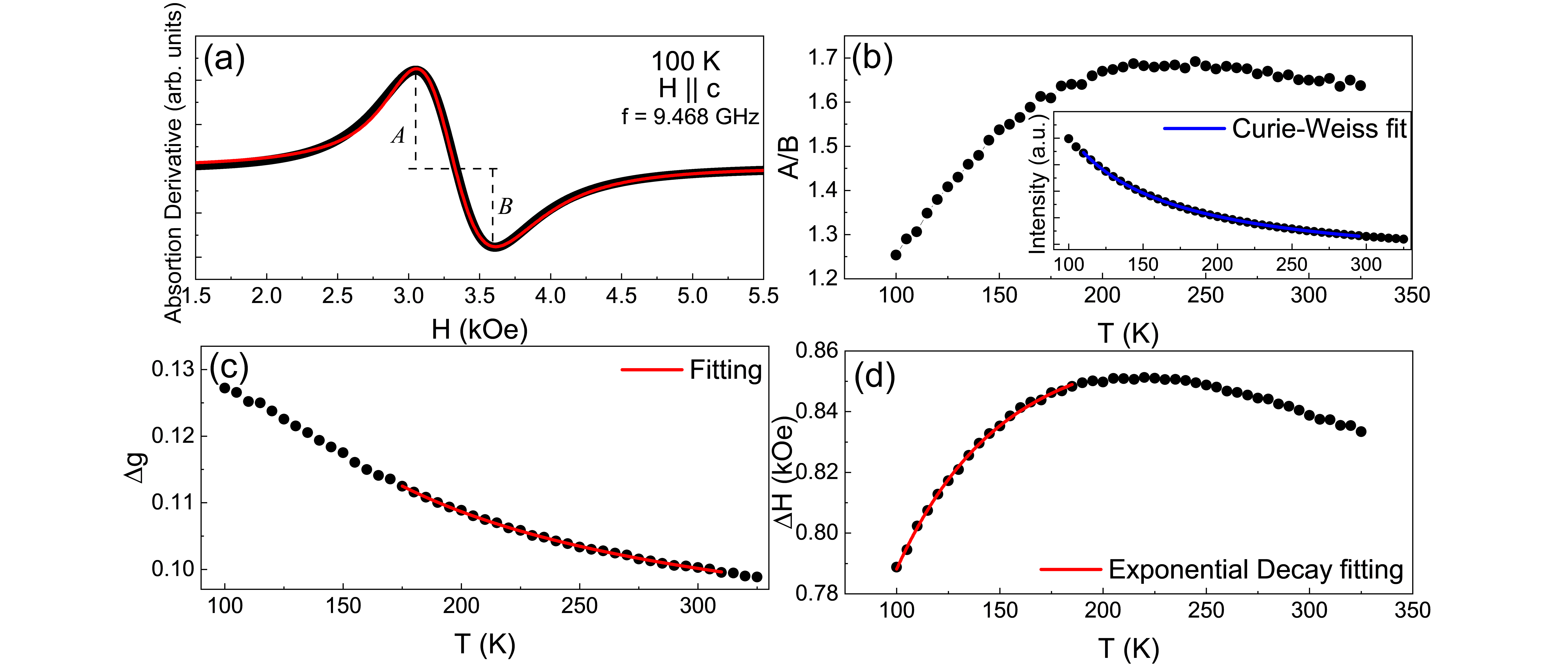}
    \end{center}

	\caption{(a) ESR spectra at \textit{T} = 100 K. (b) ESR spectra asymmetry ratio A/B (inset: ESR intensity), (c) $\Delta g$, and (d) linewidth as a function of the temperature.}
	\label{ESR}
\end{figure*}

\textbf{Electron Spin Resonance} Figure \ref{ESR}(a) shows the Eu$^{2+}$ \textit{X} band (9.47 GHz) ESR spectrum of EuZn$_{2}$P$_{2}$ with \textit{H} parallel to the \textit{c} axis measured at \textit{T} = 100 K. In the paramagnetic state, \textit{T} > 100 K, we observe Dysonian ESR lineshapes characteristic of localized magnetic moments in a lattice with a skin depth smaller than the size of the sample. The \textit{g} value and linewidth were obtained from fitting the resonances to the appropriate admixture of absorption and dispersion given by the power absorption derivative \textit{dP/dH} as a function of \textit{H}:
\begin{equation}\label{absortion}
	\dfrac{dP}{dH} \propto (1-\beta)\dfrac{d}{dx}\bigg(\dfrac{1}{1+x^{2}}\bigg) + \beta \bigg(\dfrac{x}{1+x^{2}}\bigg),
\end{equation}
where $\beta$ is the asymmetric parameter of the line shape, and $x = 2(H - H_{r})/\Delta H$, with $\Delta H$ as the line width and $H_{r}$ as the resonance field.

The asymmetry of the line shape can be estimated by the ratio $A/B$, as shown in Fig. \ref{ESR}(a).
If the spacial dimension $l$ of the sample is much smaller than the microwave skin depth $\delta$, such that $l/\delta\ll{1}$, one finds a symmetric Lorentzian ESR line shape ($A/B=1$). If the ratio $l/\delta\gg{1}$, as expected in a metallic environment, one finds an asymmetric Dysonian line shape ($A/B\approx2.7$). Thus, the ratio $A/B$ shown in Fig. \ref{ESR}(b) reveals the evolution of the microwave skin depth, which allows us to infer the conducting nature of the local environment for the Eu$^{2+}$ moments. Below 200~K, we observe a decrease in the ratio $A/B$. We note that the onset of CMR as well as the deviation from the Curie-Weiss behavior in magnetic susceptibility both occur on this temperature scale. In fact, a decrease in $A/B$ supports the formation of magnetic polarons: as free carriers are trapped into MP, an increase in the microwave skin depth should occur. Similar results have been reported for Eu$_5$In$_2$Sb$_6$, where the formation of MP is thought to occur within a similar temperature range \cite{souza_microscopic_2022}. 

The Eu$^{2+}$ ESR intensity ($I_{ESR}$) is related to the Eu$^{2+}$ spin susceptibility ($\chi_{s}$). As shown in Figure \ref{ESR}(b), the ESR intensity shows a Curie-Weiss-like behavior [$\chi_{s} \propto 1/ (T - \theta)$], wherein $\theta$ is the Weiss temperature. From a Curie-Weiss fit (red solid line), we extracted a positive $\theta$ of 33(1) K, which is a measure of the local Eu$^{2+}$-Eu$^{2+}$ short-range interaction and clearly indicates FM interactions. This value is also consistent with magnetic susceptibility data.

The experimental Eu$^{2+}$ \textit{g} value, \textit{g} = $h\nu/\mu_{B}H_{r}$, is obtained from the analysis of the Eu$^{2+}$ ESR spectra using Eq. \ref{absortion}. Here, \textit{h} is the Planck constant, and $\mu_{B}$ is the Bohr magneton. Figure \ref{ESR}(c), shows $\Delta g = g_{exp} - g_{theor}$ as a function of the temperature, wherein $g_{theor} = 1.993$. $\Delta g(T)$ shows a Curie-Weiss-like behavior related to the exchange interaction between the localized 4\textit{f} electron spin (Eu$^{2+}$) and the conduction electron spin. The observed $\Delta g(T)$ increase on cooling reveals the development of Eu$^{2+}$-Eu$^{2+}$ ferromagnetic (FM) interactions that reduce the resonance field. 

When bottleneck and dynamic effects are not present, the ESR \textit{g} shift (Knight shift) and the thermal broadening of the linewidth (Korringa rate) can be written as:
\begin{equation}\label{deltag}
	\Delta g = J_{fs}(\textbf{0}) \dfrac{\eta(E_{F})}{1-\alpha}.
\end{equation}
Here, $J_{fs}(\textbf{0})$ is the effective exchange interaction between the Eu$^{2+}$ local moment and the conduction electron in the absence of momentum transfer, $\eta(E_{F})$ the density of states for one spin direction at the Fermi surface, and $(1-\alpha)^{-1}$ is the Stoner enhancement factor that can be obtained by comparing the Pauli susceptibility with $\chi_0$ of the Curie-Weiss fit of the magnetic susceptibility. We have used a density of states of 0.017~states/(eV~mol-spin) obtained from a carrier density of $5.6\times10^{17}$ cm$^{-3}$ (from AC Hall measurements) to calculate the theoretical value of the Pauli susceptibility of 2.22 $\times10^{-6}$ emu/(mol Oe). Comparing that value with $\chi_0$ obtained from Curie-Weiss fits to the magnetic susceptibility, we extracted a value of $\alpha$ = 0.93(6). Using these obtained values and $\Delta g_{0}$ in Eq.~\ref{deltag}, we have found a $J_{fs}(\textbf{0})$ value of 350(300) meV. This relatively large value of $J_{fs}(\textbf{0})$ suggests a strong exchange coupling between the Eu$^{2+}$ local moment and the conduction electrons consistent with a magnetic polaron scenario.

The Eu$^{2+}$ ESR linewidth, $\Delta H$, shown in Figure \ref{ESR}(d), provides information on the local spin dynamics of the Eu$^{2+}$ resonant spins. The intriguing non-monotonic behavior displayed in Fig. \ref{ESR}(d) is very distinct from the linear thermal broadening of the linewidth (Korringa rate) typically observed for ESR of dilute local moments in metals \cite{PhysRevB.55.1016,PhysRevB.56.8933,PhysRevB.63.092406,PhysRevB.86.165131}. The initial broadening of the Eu$^{2+}$ ESR linewidth indicates that, even at high temperatures, the relaxation in this material is dominated by Eu$^{2+}$ spin-spin interactions, similar to other related low-carrier density materials, such as EuIn$_2$As$_2$ \cite{Rosa2012}. However, the observed narrowing for \textit{T} < 200 K may suggest a crossover to a regime dominated by the exchange narrowing effects associated with the formation of magnetic polarons, which leads to a more coherent Eu$^{2+}$-Eu$^{2+}$ spin-spin scattering.

\begin{figure*}[ht]
    \begin{center}
    \includegraphics[width=0.9\textwidth]{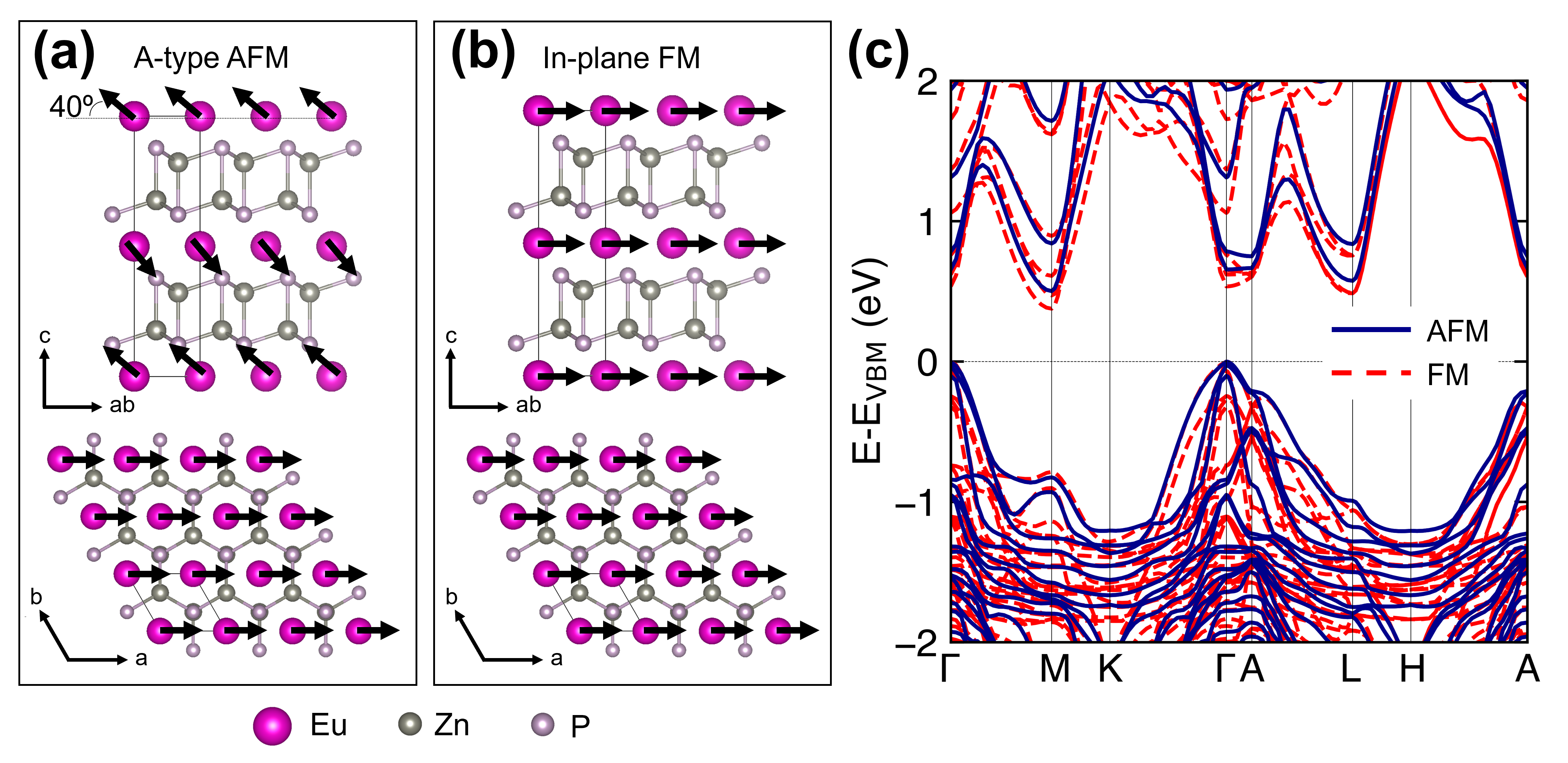}
    \end{center}
    \caption{\label{fig:DFTgroup} (a) A-type AFM ordering and (b) FM ordering (fully polarized moments under applied field) of Eu$^{2+}$ moments, which are indicated by black arrows. (c) Electronic band structure of EuZn$_2$P$_2$ for both A-type AFM and FM spin structures. The zero of energy has been shifted to the valence band maximum energy.}
\end{figure*}


\textbf{Theoretical Calculations}

The band structures for the AFM and FM orderings are shown in Fig.~\ref{fig:DFTgroup}(c). As in-plane and out-of-plane FM spin configurations yield virtually identical band structures, only the band structure for out-of-plane FM ordering is displayed. The A-type AFM ordering (Fig.~\ref{fig:DFTgroup}(a)) has an indirect band gap of 502 meV and a direct band gap of 656 meV. The FM ordering (Fig.~\ref{fig:DFTgroup}(b)) has an indirect band gap of 377 meV and a direct band gap of 536 meV, amounting to a 25\% and 18\% reduction, respectively. This band gap reduction is consistent with the experimentally observed relationship between magnetoresistance and applied field strength; namely, a reduction in the band gap occurs as the magnetic ordering transitions from the ground state A-type AFM (no applied field) to FM (fully saturated magnetization under applied field).

We note that these band gap values are substantially larger than those observed via transport in this work and those previously reported in DFT calculations using the local density approximation (LDA) \cite{krebber_colossal_2023}. In Ref. \cite{krebber_colossal_2023}, the authors implement an LDA+U scheme, selecting their Hubbard U for the Eu ions by identifying a U value that places the filled Eu bands an appropriate distance below the Fermi energy, ~1.5 eV, in agreement with ARPES measurements. In this work we take the same approach, selecting U = 6.0 eV for the Eu ions. However, as our calculations utilize PBE+U rather than LDA+U, our calculated band gaps are significantly overestimated. We emphasize that while the calculated band gaps do not quantitatively agree with experiment or prior first-principles calculations, the qualitative observation of a band gap reduction as a function of magnetic ordering should be robust to variations in the choice of exchange-correlation functional between PBE and LDA.


\section{IV. Discussion and Summary}

In summary, we have synthesized single crystals of EuZn\textsubscript{2}P\textsubscript{2} using In flux and confirmed the trigonal crystal structure using single crystal x-ray diffraction. Our transmission electron microscopy measurements do not show evidence for Zn vacancies, In interstitials, or substitutions. We characterized our samples with electrical transport, magnetization, dilatometry, and electron spin resonance measurements that collectively point to magnetic polaron formation, typical of magnetic Eu\textsuperscript{2+} systems with low carrier density. We observe an increase in electrical resistance by 7 orders of magnitude on cooling from room temperature to 2 K. These measurements reveal a cusp in the resistance at $T_N$ followed by a marked decrease below the transition. Our electrical resistance results are in contrast with those Ref. \cite{krebber_colossal_2023}, and these differences may be due to in-gap states or parallel conduction paths from flux inclusions. Though we have found no experimental evidence of inclusions or vacancies in our samples, low concentrations of these defects may go undetected. A recent review on property dependence of growth conditions in the EuT\textsubscript{2}X\textsubscript{2} ($T =$ Zn, Cd; $X =$ P, As, Sb) family show that these materials are sensitive to vacancies and impurities \cite{kliemt_chemical_2024}, thus we cannot completely rule out these effects.     

We have shown that the estimated energy gap based on electrical transport data is smaller than the direct gap (~600 meV) found using M-EELS and can vary greatly in our samples. This observed sample dependence is likely due to flux inclusions that would not affect M-EELS spectra if the measured surface is free of indium. The spectra were stable across days of measurements and multiple samples, ruling out the possibility of charge buildup from the impinging electron beam. M-EELS in the optical limit ($q\sim0$) is consistent with optical measurements \cite{chen2024}. Differences between the gap directly visualized in M-EELS and interpreted from FTIR data may be the result of the fit used for the optical data, which has been shown to be model-dependent in semiconductors \cite{zanatta2019}.

The magnetoresistance in EuZn\textsubscript{2}P\textsubscript{2} has several interesting features that have been observed in other Eu-based semiconductors as well as other members of the trigonal EuT\textsubscript{2}X\textsubscript{2} family. The magnitude of the $MR$ near the magnetic ordering temperature (-99.7\%) is of the same order as the largest reported within this family of materials. The CMR in EuZn\textsubscript{2}P\textsubscript{2} is consistent with magnetic polaron formation, which begins as high as 200~K, below where we observe a deviation from Curie-Weiss behavior (T = 150 K) as well as a continuous narrowing of ESR linewidth $\Delta{H}$ as the temperature is lowered and the size of ferromagnetic clusters grow.
ESR also finds a strong exchange interaction between Eu\textsuperscript{2+} moments and conduction electrons which is a basic requirement for magnetic polaron formation.

The peak in resistance that coincides with $T_N$ in zero field, as well as the field dependence of this peak, is very reminiscent of other Eu-based magnets where magnetic polarons are thought to form, such as EuB\textsubscript{6} \cite{glushkov_magnetic_2007}, Eu\textsubscript{5}In\textsubscript{2}Sb\textsubscript{6} \cite{rosa_colossal_2020, ale_crivillero_magnetic_2023}, and more recently EuCd\textsubscript{2}P\textsubscript{2} \cite{sunko_spin-carrier_2023, usachov_magnetism_2024}. A common aspect of the transport in these materials as well as EuZn\textsubscript{2}P\textsubscript{2} is that the CMR peaks in proximity to the magnetic ordering temperature as a partial or complete delocalization of carriers occurs close to the magnetic phase. Similar to EuZn\textsubscript{2}P\textsubscript{2}, the peak in resistivity in all of these compounds decreases in magnitude, broadens, and is pushed to higher temperature while increasing magnetic field. A sufficient field should completely suppress the peak in resistivity, as is observed in EuZn\textsubscript{2}P\textsubscript{2} at 9 T.

Interestingly, the magnitude of negative MR in EuZn\textsubscript{2}P\textsubscript{2} decreases below $T_N$ and then reaches its maximum value at the lowest temperature measured. The largest MR response occurs as the moments are polarized in field. The field values at which the MR (Fig.~\ref{fig:Resist_2panel}(c)) saturate correspond well to the field values where the moments reach a fully polarized state \cite{berry_-type_2022}. This can also be observed in EuSn\textsubscript{2}As\textsubscript{2} \cite{chen_negative_2020} and EuCd\textsubscript{2}As\textsubscript{2} under pressure \cite{du_consecutive_2022}. This effect may be due to a reduction in magnetic scattering in the polarized state, a reduction of the band gap in the polarized state, the persistence of magnetic polarons within the ordered state, or a combination of these effects. 

The lack of response in resistivity to applied uniaxial stress is in contrast to Eu\textsubscript{5}In\textsubscript{2}Sb\textsubscript{6}, where the percolation of magnetic polarons is thought to drive colossal piezoresistance in analogy to magnetic field. As Ref. \cite{PhysRevB.110.014421} has demonstrated, the resistivity and CMR in EuZn\textsubscript{2}P\textsubscript{2} changes significantly under very large hydrostatic pressures. Here, we were only able to reach a maximum of 0.6 GPa of uniaxial stress; thus more work is required to fully understand this behavior. Larger samples devoid of flux inclusions would allow for more accurate measurements along all crystallographic axes as well as allow for greater applied stress.  Differences in carrier concentration, compressibility, and anisotropy of ferromagnetic clusters could be a source of the discrepancy between the response in both materials.  Dilatometry measurements are consistent with a reorientation of AFM domains in low fields, as reported in \cite{krebber_colossal_2023}. The magnitude of the relative length change in field at 9 T has a maximum that coincides well with the peak in CMR, similar to what was observed for EuB\textsubscript{6} \cite{manna_lattice_2014}. 

In conclusion, the comprehensive set of physical property measurements we present here as well as microscopic evidence from ESR measurements are consistent with the formation of ferromagnetic clusters in EuZn\textsubscript{2}P\textsubscript{2}. Collective behavior that fit within the magnetic polaron picture seems ubiquitous in magnetic Eu\textsuperscript{2+} systems with low carrier density. Strong exchange coupling of local moments to conduction electrons inherent to these systems drives colossal magnetoresistance in these materials.


\section{Acknowledgements}
Work at Los Alamos National Laboratory was performed under the auspices of the U.S. Department of Energy, Office of Basic Energy Sciences, Division of Materials Science and Engineering. Scanning electron microscope and energy dispersive x-ray measurements were performed at the Center for Integrated Nanotechnologies, an Office of Science User Facility operated for the U.S. Department of Energy (DOE) Office of Science. Los Alamos National Laboratory, an affirmative action equal opportunity employer, is managed by Triad National Security, LLC for the U.S. Department of Energy’s NNSA, under contract 89233218CNA000001. Work at the Molecular Foundry was supported by the Office of Science, Office of Basic Energy Sciences, of the U.S. Department of Energy under Contract No. DE-AC02-05CH11231. CSK gratefully acknowledges the support of the U.S. Department of Energy through the LANL/LDRD Program and the G. T. Seaborg Institute. JS acknowledges support from the Los Alamos Institute for Materials Science. MSC acknowledges funding from the Laboratory Directed Research \& Development Program.

\appendix

\section{Appendix A: Experimental Methods}

Single crystals of EuZn\textsubscript{2}P\textsubscript{2} were grown using In flux. We opted for indium, instead of the previously-used tin, because of concerns with interstitial substitutions or doping at the Zn site. Because indium's atomic radii is larger than tin's, substitutions are less likely.  The optimal starting ratio of constituent elements is 1:2:2:30~=~Eu:Zn:P:In. This growth technique produces thin hexagonal plates with large facets that are normal to the c axis. Flux on the surfaces of the samples was removed mechanically. In flux inclusions are fairly common in the bulk and can be clearly seen when cutting or polishing the samples. 

X-ray diffraction experiments were performed at room temperature using a Bruker D8 Venture single crystal diffractometer with an Incoatec I$\mu$S microfocus source (Mo K-$\alpha$ radiation, $\lambda = 0.71073$ \r{A}) as the radiation source. Data were collected using a PHOTON II CPAD area detector. Raw data were processed with Bruker SAINT software, including multi-scan absorption correction. The initial crystallographic model was obtained via the intrinsic phasing method in SHELXT. A least-squares refinement was performed using SHELXL2018. 

HR-STEM images were acquired on the TEAM I, a customized FEI Titan, with a direct-electron K3 detector, operated at 300 keV with a 20 mrad semi-convergence angle. HR-TEM images were collected on a Titan Environmental TEM with a K3 detector, operated at 300 keV in bright-field mode.

M-EELS measurements were taken at 300 K using a beam energy of 50 eV, with a full-width at half-maximum of $<10$ meV. Experiments were carried out along the $(H,0)$ direction. 
The samples were cleaved along the (00L) direction \textit{in-situ} in vacuum better than $3\times10^{-10}$ Torr.

Magnetization measurements were performed in a commercial MPMS SQUID magnetometer. Resistance was measured with a DC 2-wire method using a Keithley 2400 Sourcemeter with 1 V of constant bias. Resistivity measurements under uniaxial pressure were performed with a commercial pressure cell (Razorbill FC-100) with current and pressure along the same direction. The AC Hall effect was measured on a standard PPMS puck with a custom coil attached to the surface that is able to produce the required AC magnetic field. These measurements were taken at several magnetic field frequencies to check for capacitive contributions to the AC Hall voltage. Current was applied to the sample using a Keithley 2400 while the in phase and out of phase voltages were detected using a Stanford Research SR860 lock-in amplifier. A capacitive dilatometer as described in Ref. \cite{schmiedeshoff_versatile_2006} was used for thermal expansion and magnetostriction measurements.

{\renewcommand{\arraystretch}{1.3}
\begin{table*}[!ht]
\caption{Fractional atomic coordinates and isotropic displacement parameters of the refinement of EuZn$_2$P$_2$ with fixed occupations.}
\label{tab:atomic_sites_Zn_fixed_varied_occ}
\begin{tabular}{*8c}
\hhline{========}

 Atom & Wyckoff & x   & y   & \multicolumn{2}{c}{z}    & \multicolumn{2}{c}{U$_{\mathrm{iso}}$ (\r{A}$^2$) }     \\ \hline
  & & & & Fixed & Varied & Fixed & Varied \\
 \hline
Eu & 1a & 0   & 0   & 0    & 0       & 0.00754(19)      & 0.00741(19)   \\ 
Zn & 2d & 2/3 & 1/3 & 0.36947(10) & 36952(10) & 0.0093(2) & 0.0088(3)  \\ 
P  & 2d & 1/3 & 2/3 & 0.26913(18) & 0.26914(17) & 0.0066(3)  & 0.0064(3)   \\

\hhline{========}
\end{tabular}
\end{table*}
}

{\renewcommand{\arraystretch}{1.3}
\begin{table*}[!ht]
\caption{Anisotropic displacement parameters of EuZn$_2$P$_2$ with fixed occupations. U$_{23}$ and U$_{13}$ are 0 by symmetry.}
\label{tab:ADPs_fixed_Zn_occ}
\begin{tabularx}{\textwidth}{XXXXX}
\hhline{=====}
Atom & U$_{11}$ (\r{A}$^2$)   & U$_{22}$  (\r{A}$^2$)  & U$_{33}$  (\r{A}$^2$)  & U$_{12}$  (\r{A}$^2$) \\ \hline
Eu & 0.0077(2) & 0.0077(2)   & 0.0072(2)   & 0.00385(10)  \\ 
Zn & 0.0093(3) & 0.0093(3) & 0.0093(4) & 0.00466(13)  \\ 
P  & 0.0070(4) & 0.0070(4) & 0.0060(6) & 0.00348(18)  \\ \hhline{=====}
\end{tabularx}
\end{table*}
}

{\renewcommand{\arraystretch}{1.3}
\begin{table*}[!ht]
\caption{Anisotropic displacement parameters of EuZn$_2$P$_2$ with Zn occupation varied. U$_{23}$ and U$_{13}$ are 0 by symmetry.}
\label{tab:ADPs_varied_Zn_occ}
\begin{tabularx}{\textwidth}{XXXXX}
\hhline{=====}
Atom & U$_{11}$  (\r{A}$^2$) & U$_{22}$  (\r{A}$^2$)  & U$_{33}$  (\r{A}$^2$)  & U$_{12}$ (\r{A}$^2$)  \\ \hline
Eu & 0.0076(2) & 0.0076(2)   & 0.0071(2)   & 0.00379(10)  \\ 
Zn & 0.0089(3) & 0.0089(3) & 0.0087(4) & 0.00445(15)  \\ 
P  & 0.0067(4) & 0.0067(4) & 0.0058(5) & 0.00337(18)  \\ \hhline{=====}
\end{tabularx}
\end{table*}
}

DFT calculations were performed with a plane-wave basis and projector augmented wave (PAW) pseudopotentials \cite{kresse_ultrasoft_1999} as implemented in the Vienna ab initio simulation package (VASP) \cite{kresse_efficiency_1996,kresse_efficient_1996}. Crystal relaxation and band structure calculations were performed in the generalized gradient approximation (GGA) as implemented by Perdew, Burke, and Ernzerhof (PBE) \cite{perdew_generalized_1996}. Localization of the Eu \textit{f} electrons was addressed by the addition of a Hubbard U correction, with U = 6.0 eV, using the method implemented by Dudarev, et al. \cite{dudarev_electron-energy-loss_1998}. The crystal structure of pristine EuZn\textsubscript{2}P\textsubscript{2} was relaxed with the lattice parameters fixed to the experimental values of a = 4.087 Å and c = 7.010 Å \cite{klufers_ab_1980}. The internal coordinates of a 1x1x2 supercell were allowed to relax using a plane-wave energy cutoff of 600 eV, a 9x9x3 Gamma-centered k-grid, and collinear A-type AFM ordering (alternating between planes along the c-axis) until forces converged to < 1 meV/Å. The band structure was calculated with a reduced energy cut-off of 500 eV and the addition of spin-orbit coupling (SOC) using the canted A-type AFM ordering reported by Krebber \textit{et al.} (Fig.~\ref{fig:DFTgroup}(a)) \cite{krebber_colossal_2023}. The band structure was additionally calculated using FM ordering (Fig.~\ref{fig:DFTgroup}(b)) oriented both in-plane (parallel to the a axis) and out-of-plane (parallel to the c axis).

\section{Appendix B: Crystal Structure Refinement Tables}

To check for disorder on the Zn site, the refinement procedure was performed twice; once with all site occupancies fixed and again letting just the Zn site occupancy vary. The metrics of each refinement and the refined Zn occupancy are shown in Table \ref{tab:refinements}. When Zn was allowed to vary, the occupancy remained close to 100 percent, but there was a slight improvement in the overall residual, weighted residual, and the goodness of fit. Since Eu has a Wyckoff position of 1a, it was not permitted to shift between the two refinements. However, it did show a modest decrease in the isotropic atomic displacement parameter (ADP). See Table \ref{tab:atomic_sites_Zn_fixed_varied_occ}. The z-position of the Zn and P atomic sites changed slightly, along with their isotropic ADPs, seen in Table \ref{tab:atomic_sites_Zn_fixed_varied_occ}. The most significant decrease is observed for the Zn ADPs, with the largest decrease being in the U$_{33}$ parameter. See Tables \ref{tab:ADPs_fixed_Zn_occ} and \ref{tab:ADPs_varied_Zn_occ}.

To summarize, though there is an improvement in the refinement when the Zn site occupation is allowed to vary, the effect is modest. The refined occupancy of 99.2(3)$\%$ would imply a vacancy density of $\sim 9.9 \times 10 ^{-5}$ \r{A}$^{-3}$. This is consistent with there being no indium flux incorporated into the lattice, which is also supported by TEM/STEM results.

\section{Appendix C: In Substitution on Zn Sites}

\begin{figure}[!ht]
\includegraphics[width=3.375in]{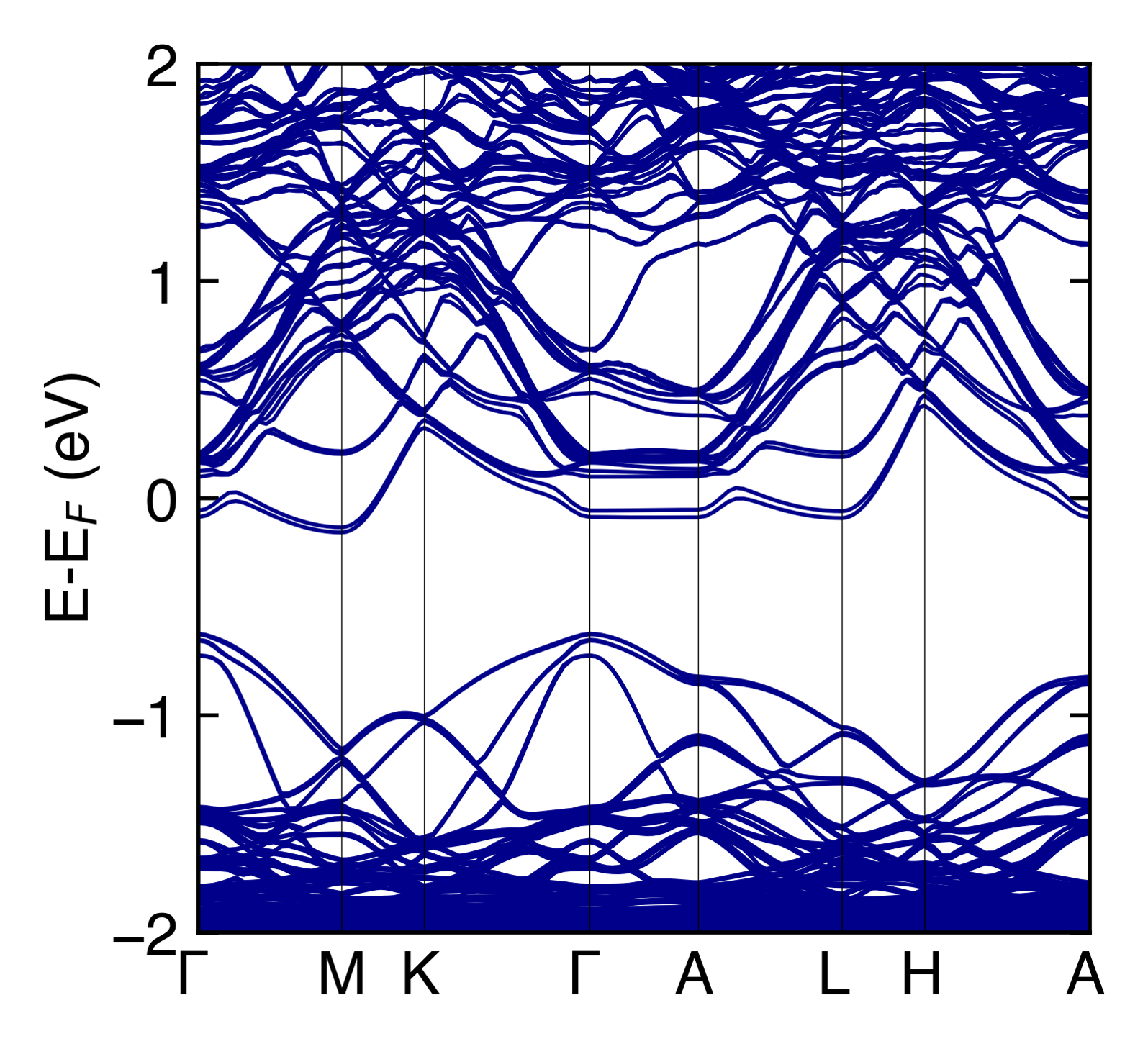}
\caption{\label{fig:appendft} Electronic band structure of EuZn\textsubscript{2}P\textsubscript{2} with 3\% substitution of In on Zn sites.}
\end{figure}

As the transport measurements of EuZn\textsubscript{2}P\textsubscript{2} show sample dependence, we calculate the DFT electronic structure of a 3x3x2 EuZn\textsubscript{2}P\textsubscript{2} supercell with ~3\% substitution of In on Zn sites, using the same exchange-correlation and Hubbard U corrections as described in the main text. The lattice parameters were fixed and the crystal structure was relaxed using a 500 eV energy cutoff, a 3x3x3 Gamma-centered k-grid, A-type AFM collinear magnetism, and forces converged to < 1 meV/Å. The band structure was calculated with spin-orbit coupling (SOC) using the canted A-type AFM ordering reported in Ref.~\cite{krebber_colossal_2023}. 

The calculated band structure is shown in Fig.~\ref{fig:appendft}. The primary effect of In substitution on the band structure is the appearance of a partially filled band near the conduction band edge of predominant In character. As this band is partially filled, it shifts the Fermi energy to higher energy, closer to the conduction band edge. With ~3\% substitution of In on Zn sites the band structure is metallic, which is not consistent with the experimental transport measurements that indicate that even with defects or flux inclusions, the EuZn\textsubscript{2}P\textsubscript{2} sample remains gapped. However, it is reasonable to conclude that, extrapolated to more dilute concentrations, the presence of substitutional In defects on Zn sites could result in localized in-gap states near the conduction band edge, behaving like an n-type dopant. While we have not uncovered conclusive evidence of presence of substitutional defects in our samples, the clear presence of In flux inclusions suggests that a small amount of In substitutions on Zn sites could potentially be present; this could contribute, in part, to the variability in the band gap across different regions of the sample and the measurement of much smaller band gaps than those reported in previous literature. 

\section{Appendix D: Hall coefficient from AC Hall effect}

The contributions to the experimental Hall voltage $V_e$ in a typical DC field measurement are given by,
\begin{eqnarray}
V_e=\frac{R_HIB}{t}+V_m+V_{TE},
\label{eq:DChall1}
\end{eqnarray}
where $R_H$ is the Hall coefficient, $I$ is the current, $B$ is the applied magnetic field, $t$ is the thickness of the sample, $V_m={\alpha}{\rho}I/t$ is the misalignment voltage with $\alpha$ as a geometric factor between 0 and 1, and $V_{TE}$ is the thermoelectric voltage that depends on the thermal gradients present \cite{lindemuth_hall_2011}. The misalignment voltage depends on the resistivity of the measured materials, thus in insulating materials this contribution is large. 

To mitigate these effects, a sample can be placed within a low frequency alternating magnetic field of frequency $\omega$ with current applied using a typical Hall measurement configuration. The Hall voltage will then have the form,
\begin{eqnarray}
V_e=\frac{R_HIBcos(\omega{t})}{t}+{\beta{B}{\omega}cos(\omega{t})}+V_m+V_{TE},
\label{eq:DChall2}
\end{eqnarray}
where the second term is due to a finite inductive voltage from the sample and sample wiring. As the thermoelectric and offset voltage contributions do not depend on field, they are now separated in frequency space. Through the use of lock-in amplifier detection and current reversal to remove the inductive voltage contribution, the Hall coefficient can be solved for as, 
\begin{eqnarray}
R_H=\frac{(V_{AC}(I_+)-V_{AC}(I_-))t}{Bcos(\omega{t})(I_+-I_-)}.
\label{eq:DChall3}
\end{eqnarray}


\bibliography{bib_001.bib}

\end{document}